\documentclass[12pt, letterpaper]{article}
\usepackage[margin=2cm]{geometry}

\usepackage{amssymb}
\usepackage{amsmath}
\usepackage{graphicx}
\usepackage{physics}

\newcommand{\Gr}{\mathsf{G}}
\usepackage{cite}
\usepackage{url}
\newcommand{\at}[2][]{#1\bigg|_{#2}}

\usepackage{nicefrac}   
\newcommand{\half}[1]{\nicefrac{#1}{2}}
\newcommand{\third}[1]{\nicefrac{#1}{3}}
\newcommand{\nfrac}[2]{\nicefrac{#1}{#2}}
\newcommand{\Li}[1]{\text{Li}_{#1}}

\usepackage{color}

\newcommand{\tableheight}{\textcolor{white}{$\lim\limits_{x\to 1^-} x \ $}}

\title{ Bose-Einstein statistics for a finite number of particles.}

\author{ Pedro Pessoa$^1$  \\
$^1$Department of Physics, University at Albany - SUNY \\  Albany, NY - USA}
\date{}

\begin{document}

\maketitle
\abstract{
This article presents a study of the grand canonical Bose-Einstein (BE) statistics for a finite number of particles in an arbitrary quantum system. 
The thermodynamical quantities that identify BE condensation --- namely, the fraction of particles in the ground state and the specific heat --- are calculated here exactly in terms of temperature and fugacity. These calculations are complemented by a numerical calculation  of fugacity in terms of the number of particles, without taking the {thermodynamic} limit. 
The main advantage of this approach is that it does not rely on approximations made in the vicinity of the usually defined critical temperature, rather it makes calculations with arbitrary precision possible, irrespective of temperature. 
Graphs for the calculated thermodynamical quantities are presented in comparison to the results previously obtained in the {thermodynamic} limit.
In particular, it is observed that for the gas trapped in a 3-dimensional box the derivative of specific heat reaches smaller values than what was expected in the {thermodynamic} limit --- here, this result is also verified with analytical calculations. 
This is an important result for understanding the role of the {thermodynamic} limit in phase transitions and makes possible to further study BE statistics without relying neither on the {thermodynamic} limit nor on approximations near critical temperature. 
}


\newpage

\section{Introduction}
\paragraph{}
In most textbooks on statistical physics, e.g \cite{Landau,Robertson93,Salinas00}, Bose-Einstein (BE) condensation is taught as one of the first examples of phase transitions. 
This phase transition is identified by the fact that there is a low value in temperature, namely critical temperature, for which the  specific heat for a non-interacting quantum gas of bosons is not analytical. 
Physically the condensation/phase transition happens because a thermodynamically relevant fraction of particles inhabits the quantum state of lowest energy, namely the ground state. The fraction of particles in the ground state itself also exhibits a non-analytical behaviour at critical temperature, indicative of phase transition.

BE condensation was observed experimentally in 1995 \cite{Anderson95,Bradley95,Davis95}. In these experiments the number of particles ranged in the order of thousands \cite{Anderson95} to millions\cite{Davis95}. This sprung interest in studies of BE statistics that do not rely on the thermodynamic limit (number of particles $N \to \infty$) \cite{Ketterle96,Napolitano97,Pathria98,Elivanov00,Pham18,Serhan21,Jaouadi11,Noronha13}. It is important to notice that the non-analytical behaviour mentioned previously is a consequence of the thermodynamic limit applied to BE statistics and therefore shouldn't be expected for a finite number of particles.

In the present article I obtain more general results allowing for the studies of a broader range of quantum systems. Specifically, the present article studies quantum systems for which the density of states is proportional to a power of the energy\footnote{This relationship between the density of states and energy is known as dispersion relations. For further studies that investigate BE statistics for general dispersion relations see e.g. \cite{Aguilera-Navarro99,Li99,Pessoa21}}, $\Gr(\epsilon) \propto \epsilon^\eta$. A $D$-dimensional harmonically trapped gas is identified with $\eta = D-1$, while the well-studied example of a gas in a box is identified with $\eta = \half{D}-1$.

The major difficulty halting progress in this investigation comes from the lack of a closed form expression for the fugacity $\xi$ of a BE gas in terms of the number of particles $N$ and temperature $T$, which is given by 

\begin{equation}\label{NBE}
    N =  \kappa \frac{ \Gamma(\eta+1)}{\beta^{\eta+1}} \Li{\eta+1}( \xi)   + \frac{\xi}{1-\xi} \ ,
\end{equation}
where $\beta$ is the unit corrected inverse temperature, $\beta = \frac{1}{k_{B}T}$ where $k_{B}$ is the  Boltzmann constant, and $\Li{}$ refers to the polylogarithm family of functions  \cite{wolframpages}

\begin{equation} \label{Polylogdef}
\Li{\varphi}(y) = \frac{1}{\Gamma(\varphi)} \int_0^\infty du \ \frac{u^{\varphi-1}}{y^{-1} e^u-1} = \sum_{k=1}^\infty \frac{y^k}{k^\varphi}\ ,
\end{equation}
with $\Gamma(\varphi)$ being the Euler's gamma function. BE statistics assumes $\xi \in [0,1)$. 
Note that for fixed $\beta$ and $\eta>0$, $N$ is strictly increasing with $\xi$. Therefore $\xi(\beta,N)$ is well-defined as the inverse of \eqref{NBE}. This, however, has not been written in closed analytical form to the best of my knowledge\footnote{Progress has been made\cite{Kim19} by defining polyexponential functions as the inverse of \eqref{Polylogdef}. However, to the best of my knowledge, these polyexponentals were only written in closed analytical form for integer values of $\varphi$. \label{f1} }. 
This is fundamental for the study of BE statistics for a finite number of particles since the thermodynamical quantities that identify the phase transition --- fraction of particles in the ground state and specific heat --- can be written exactly in terms of $\beta$ and $\xi$, as we will see in the present article.

Most of the investigations done in BE statistics for finite number of particles \cite{Ketterle96,Napolitano97,Pathria98,Elivanov00,Pham18,Serhan21} is restricted to harmonically trapped gases, $\eta = D-1$.
Among the investigations that went beyond the study of harmonically trapped gases, it is important to mention the work of Jaouadi et. al.\cite{Jaouadi11} --- that investigated the 3 dimensional gas of bosons in a power-law trap --- and the work of  Noronha \cite{Noronha13} --- investigating the statistics for the Bose gas for a particular set of spaces and potentials, namely the particle in a box and a three-sphere. 
Both of these works, however, rely on approximations taken in the vicinity of critical temperature. 

The present article differs from these aforementioned works since it does not assume a specific quantum system, rather here I presented an exact calculation of thermodynamical quantities in terms of $\beta$ and $\xi$ for arbitrary $\eta$ --- later will be explained that one can only identify condensation for $\eta>1$ --- making it general. These calculations are presented for the first time here, to the best of my knowledge. Moreover, these exact results are complemented by the numerical calculation of $\xi(\beta,N)$, so all thermodynamical quantities can be calculated for a finite number of particles, allowing for reliable results irrespective of temperature.

A numerical implementation of $\xi(\beta,N)$ is found
in my GitHub repository \cite{github}.  It is important to point out that the calculation of fugacity can not be simply implemented with the floating point arithmetic built in most computer languages, since for large $N$, $\xi$ is extremely close to unity. This difficulty is avoided by using the mpmath python library \cite{mpmath} which allows for arbitrary precision float point arithmetic --- roughly speaking, all quantities can be calculated with an arbitrary number of decimal places\footnote{The graphs presented here are based on a calculation with 50 decimal places.}.

From the fugacity obtained numerically, one can calculate the fraction of particles in the ground state, specific heat and specific heat first derivative and compare to the results obtained for them in the thermodynamic limit, available in \cite{Landau,Robertson93,Salinas00,Aguilera-Navarro99}. For transparent presentation of the results, graphs for all the calculated quantities are presented for number of particles ranging from $N= 10^2$ to $N= 10^7$, and its comparison to the thermodynamic limit.

The layout of the present article is as follows: Sec. \ref{spectrum} will review the statistical mechanical description of quantum gases in terms of $\beta$ and $\xi$ and obtain the parameter $\eta$ for some well known models.  
Sec. \ref{limitsection} will present the description of Bose gases in the thermodynamic limit for different values of $\eta$ and show how these values affect the convergence of critical temperature and the existence of phase transition identified by non analytical behaviour.
Sec. \ref{finito} presents exact results for the fraction of particles in the ground state and specific heat in terms of $\beta$ and $\xi$. These quantities are graphed  in terms of $\beta$ and $N$ from the numerical implementation of $\xi(\beta,N)$ found in \cite{github}. These calculations do not specify a value of $\eta$ and the graphs are presented from $\eta=\half{1}$ and $\eta=2$.

\section{Spectrum and density of states} \label{spectrum}
\paragraph{}
Quantum statistical mechanics consists of assigning probability distributions for the space of occupancy number of each quantum state --- also referred to as Fock space. Mathematically this means to assign a probability $\rho(x)$ where $x=\{x_i\}$ in which $i$ enumerates the quantum states. In this description, $x_i$ is the number of particles occupying the state $i$ --- for bosons $x_i$ takes positive integer values, $x_i\in [0,1,2,\ldots )$ --- and each quantum state is associated to an energy value $\epsilon_i$. 
The relationship between $i$ and $\epsilon_i$  is also referred to as the spectrum and is obtained through regular methods in quantum mechanics \cite{Sakurai}, that means $\varepsilon_i$ are the eigenvalues of the Hamiltonian operator $\hat{H}$ defined as

\begin{equation}
    \hat{H} \doteq \frac{\hat{p}^2}{2m}  + \phi(q) \ ,
\end{equation}
where $q = \{q_1,q_2,\ldots q_D\}$ are the system's position coordinates --- which, for simplicity, are assumed to be rectangular --- $\hat{p}$ is the  momentum  identified as {$\hat{p} = -i\hbar\sum_{\mu=1}^D \textbf{e}_\mu \pdv{}{q_\mu}$ --- where $\textbf{e}_\mu $ refers to the unit vector in the direction of the $\mu$-th dimension ---} and  $\phi(q)$ is the potential, $m$ refers to the mass of the confined particles, and $\hbar$ is the reduced Planck constant. For mathematical simplicity, one can shift the energy spectrum so that the energy of the ground state is zero, $\min_i \epsilon_i = 0$.

In the grand canonical ensemble $\rho(x)$ is assigned as the maximum entropy distribution constrained on the total number of particles, $N\doteq \sum_i \expval{x_i}$, and the system's internal energy, $U\doteq \sum_i \epsilon_i \expval{x_i}$. This leads to the grand canonical Gibbs distribution 

\begin{equation}\label{BEGibbs}
    \rho(x|\beta,\xi) \propto \prod_i e^{-\beta \epsilon_i x_i} \xi^{x_i} \ ,
\end{equation}
where, as mentioned, $\beta = \frac{1}{k_B T}$ and the fugacity $\xi$ is identified to the chemical potential $\mu$ as $\xi = e^{\beta\mu}$.
The number of particles $N$ and the internal energy $U$ can be calculated from \eqref{BEGibbs} as

\begin{subequations} \label{Averages}
\begin{align}
    N & = \sum_i \frac{\xi}{ e^{\beta  \epsilon_i } \ - \xi } \label{sumN}   \\
    U & = \sum_i \frac{\epsilon_i \ \xi }{   e^{\beta \epsilon_i }  -  \xi } \label{sumU} \ .
\end{align} 
\end{subequations}
The details of how the Gibbs distribution \eqref{BEGibbs} is derived as well as how it leads to \eqref{Averages} are presented in Appendix \ref{appendixMaxEnt}.

In order to calculate $N$ and $U$ in terms of $\beta$ and $\xi$ one needs the full spectrum of energies. I am not aware of any calculations of the series \eqref{Averages} in closed form for a general spectrum nor any calculation directly from \eqref{Averages} for physically relevant systems. 

Because of this, generally the study of quantum statistics mechanics relies on treating the spectrum as a continuous --- justified when the energy of the system is much larger than the differences of energy in the spectrum, $ U \gg \max_{i,j} |\epsilon_i-\epsilon_j|$. In such approximation the summations in \eqref{Averages} can be substituted as

\begin{equation} \label{continuous}
    \sum_i \longrightarrow \int \dd \epsilon \ \Gr(\epsilon) = \kappa \int \dd \epsilon \ \epsilon^\eta \ ,
\end{equation}
where $\kappa$ and $\eta$ are parameters to be calculated from the full spectrum of energies, as it will be presented below. Note that $\kappa$ has units of $[\text{energy}]^{(\eta+1)}$ --- or $\kappa^{\nicefrac{1}{(\eta+1)}}$ has units of energy --- one can choose a system of units in which $\kappa = 1$, or, equivalently, use $\kappa^{\nicefrac{1}{(\eta+1)}}$  as the unit of energy given by the system.

As a first example of the continuous approximation, one can study the $D$-dimensional gas trapped in a regular box of edge length $L$, related to a potential $\phi$ of the form: $\phi(q) = 0$ if $\min_\mu |q_\mu| \leq \frac{L}{2}$ and $+\infty$ otherwise. In this potential, the quantum states will be enumerated by $D$ integers --- $i = \{n_1,n_2, \ldots, n_D\}$ --- and the spectrum of energies is given by \cite{Sakurai} 

\begin{equation}
    \epsilon_i = \frac{\pi^2 \hbar^2 }{2m L^2} \sum_{\mu=0}^D n_\mu^2 \ .
\end{equation}
In this case the continuous approximation yields \cite{Robertson93, Aguilera-Navarro99, Noronha13, Chatterjee14, Pessoa21} 

\begin{equation}\label{ketaideal}
    \eta = \frac{D}{2} -1 \qq{and} \kappa= \frac{\mathsf g_s V}{\Gamma(\half D)}\left( \frac{m}{2 \pi \hbar^2}\right)^{\half{D}} \ ,
\end{equation}
where $V$ is the volume of the box, $V=L^D$, and $\mathsf g_s $ is the multiplicity of energy levels of a particle with spin $s$, $\mathsf g_s = 2s+1$. 

Another useful example is the $D$-dimensional gas in a harmonic trap, meaning a  potential of the form: $\phi(q) = \frac{m}{2} \omega^2 \sum_\mu q_\mu^2$. 
In this potential, the quantum states will be enumerated by $D$ positive integers --- $i = \{n_1,n_2, \ldots, n_D\}$ --- and the spectrum of energies is given by \cite{Sakurai} 

\begin{equation}
    \epsilon_i = \hbar \omega \sum_{\mu=0}^D n_\mu \ ,
\end{equation}
the extra term $\frac{1}{2}\hbar \omega D$ in the energy of a harmonic trapped particle is ignored, in order to assign zero energy to the ground state --- $n_\mu = 0$ for all $\mu$.
In this case the continuous approximation yields \cite{Ketterle96, Pathria98, Aguilera-Navarro99, Chatterjee14, Pessoa21} 

\begin{equation}\label{ketaharmonic}
    \eta = D-1  \qq{and} \kappa=  \frac{ \mathsf g_s}{\Gamma(D)} \left(\frac{1}{\hbar \omega}\right)^D \ .
\end{equation}

As final example --- useful to describe a more general set of quantum systems --- one can calculate the values of $\kappa$ and $\eta$ for a potential of the form $\phi(q) = \phi_0 \sum_\mu \qty|2\frac{q_\mu}{L}|^\tau$. The density of states can be calculated in a semi-classical manner\footnote{Here `semi-classical' means that it will be supposed that the density of states arising from classical mechanics is the same as the one obtained from quantum states in the continuous approximation.} from the phase space volume, meaning 

\begin{equation}
    \int_0^E \dd \epsilon \ \Gr(\epsilon) = \frac{1}{(2\pi\hbar)^D}\int_{\mathcal{A}(E)} \dd^D p \ \dd^D q \ ,
\end{equation}
where $\mathcal{A}(E) = \{ (p,q) : \frac{1}{2m} \sum_\mu p_\mu^2 + \phi(q) \leq E\}$ as presented in \cite{Li99,Wang05,Jaouadi11}. This yields

\begin{equation}\label{ketapl}
    \eta = \frac{D}{2} + \frac{D}{\tau} -1  \qq{and}   \kappa =  \frac{\mathsf g_s V}{\Gamma(\half D + \nfrac{D}{\tau})}\qty( \frac{2 m}{\pi \hbar^2})^{\half{D}} \qty[\frac{\Gamma(\nfrac{1}{\tau}+1)}{{\sqrt{2 \phi_0}}}]^D \ .
\end{equation}
It is straightforward to see that when the potential $\phi(x)$ is a harmonic oscillator --- $\tau = 2$ and $\qty(\frac{2}{L})^2 \phi_0 = \frac{m}{2} \omega^2$ --- the values of $\eta$ and $\kappa$ in \eqref{ketapl} match those of \eqref{ketaharmonic}. On the same note, when $\phi(x)$ reduces to the potential of a particle in a box --- $\tau \to \infty$ and $\phi_0 = 1$ --- the values of $\eta$ and $\kappa$ in \eqref{ketapl} match those of \eqref{ketaideal}. 

Having the values of $\eta$ and $\kappa$ from the energy spectrum, one can calculate the number of particles and the internal energy \eqref{Averages} using the continuous approximation \eqref{continuous} thus obtaining

\begin{subequations}\label{Averages4}
\begin{align}
    N &=  \kappa \frac{ \Gamma(\eta+1)}{\beta^{\eta+1}} \Li{\eta+1}(\xi) + n_0 \ , \ \qq{where} \ n_0 = \frac{\xi}{1-\xi} \ , \label{LiN}\\
    U  &=  \ \kappa \frac{ \Gamma(\eta+2)}{\beta^{\eta+2}} \Li{\eta+2}(\xi) 
     \ .\label{LiU}
\end{align}
\end{subequations}
Note that $n_0$ is the number of particles in the ground state, which needs to be added ad hoc since the continuous approximation \eqref{continuous} assigns no particle in the ground state, $\Gr(0) = 0$. It can also be seen that $n_0$ is the term in $N$ equivalent to the zero energy state in \eqref{sumN}. As will be discussed in the remainder of the present article, the addition of this term is fundamental for the study of BE condensation. 

Before studying the thermodynamical consequences of \eqref{Averages4}, two important properties of polylogarithms need attention for future use. First, it is useful to note that from \eqref{Polylogdef} one obtains

\begin{equation}
    \pdv{}{y} \Li{\varphi}(y) = \frac{1}{y} \Li{\varphi-1}(y) \ . 
\end{equation}
Second, as deduced by Cohen et. al. \cite{Cohen92}, for non-integer $\phi$ polylogarithms can be written as as a series 
\begin{equation}\label{PLseries}
    \Li{\varphi}(y) = \Gamma(1-\varphi) (-\log (y))^{\varphi-1} + \sum_{k=0}^\infty \zeta(\varphi-k) \frac{ (\log y)^k}{k!} \ ,
\end{equation}
where $\zeta$ refers to the Riemann's zeta function, $\zeta(\varphi) = \sum_{k=1}^\infty k^{-\varphi}$. This series expression is valid for $|\log y| <1 $. A similar expression of integer $\varphi$ is also found in \cite{Cohen92}.
Note that for $\varphi > 1$ it implies that $\lim\limits_{y\to1^-} \Li{\varphi}(y) =\zeta(\varphi)$ while  $\varphi \leq 1$ implies $\lim\limits_{y\to1^-} \Li{\varphi}(y) =\infty$.
These expressions will be useful in the remainder of the article.

\section{BE statistics in the thermodynamic limit}\label{limitsection}
\paragraph{}
This section will describe the phase transition for BE statistics in the thermodynamic limit by presenting the non-analytical form of the fraction of particles in the ground state and the specific heat.  Appendix \ref{appendixcv}  will show how these quantities derive from the thermodynamical quantities presented in this section follow from the calculations made for finite $N$ in Sec. \ref{finito}.  These calculations will leave $\eta$ undetermined, they reduce to those found in textbooks \cite{Landau,Robertson93,Salinas00} when $\eta=\half{1}$.

When studying BE condensation, $\beta_c$ defined as

    \begin{equation}\label{critical}
        \beta_c \doteq \left[ \kappa \frac{\Gamma(\eta+1)}{N} \lim_{\xi \to 1^-} \Li{\eta+1}(\xi)\right]^{\frac{1}{\eta+1}} \ ,
    \end{equation}
is identified as the inverse critical temperature\footnote{Other definitions for critical temperature were studied before, see e.g. \cite{Ketterle96,Noronha16,Cheng21}. These studies however are beyond the scope of the present article. }. This definition is motivated as the temperature for which $\xi$ goes to 1 when the ground state particles are ignored in \eqref{LiN}.
From \eqref{PLseries} it follows that for $\eta \leq 0$ $\beta_c$ diverges --- or the critical temperature goes to the absolute zero.  When $\eta$ is positive  $\beta_c$ converges and it yields $\beta_c = \left[ \kappa \frac{\Gamma(\eta+1)}{N} \zeta(\eta+1) \right]^{\frac{1}{\eta+1}}$. 
For the discussion presented here, one can assume $\eta>0$, guaranteeing a positive critical temperature.  
Note that this means, per Sec. \ref{spectrum}, a 1 or 2 dimensional Bose gas in a box will not have a positive critical temperature. Similarly, a 1 dimensional Bose gas in a box has divergent $\beta_c$.

A sequence of assumptions is applied when studying the Bose gas in the thermodynamic limit. These can be summarized as 
\begin{itemize}
    \item \emph{For $\beta < \beta_c$ :} treat the calculations of thermodynamical quantities as if $n_0 = 0$,
    \item \emph{For $\beta \geq \beta_c$ :} treat the calculations of thermodynamical quantities as if $\xi=1$.
\end{itemize}
This leads to a non-analytical behavior in thermodynamical quantities as will be presented below. Before presenting the results in the thermodynamic limit, they will be calculated for a finite number of particles in Sec. \ref{finito} and the thermodynamic limit will be taken in Appendix \ref{appendixcv}. It will also present comparisons in the form of graphs between the quantities calculated in thermodynamic limit  and for a finite number of particles.
 
The fraction of particle in the ground state is given by
\begin{equation}\label{tildefraction}
    \frac{\tilde{n}_0}{N} =
    \left\{\begin{array}{lll}
                0 & \text{for} & \beta<\beta_c \\
                1 - \qty(\frac{\beta}{\beta_c})^{-(\eta+1)} & \text{for} & \beta \geq \beta_c 
            \end{array}\right. \ ,
\end{equation}
for the remainder of the present article I will use the tilde notation as above to mean that the quantity is calculated in the thermodynamic limit. 

Similarly the specific heat\footnote{In many texts $c_v$ is referred to as specific heat at constant volume, as explained in Sec. \ref{spectrum}, $\kappa$ condenses the dependence with volume. That is why our derivatives in \eqref{cvlim} and \eqref{diffcvlim} are taken under constant $\kappa$. } defined as $c_v \doteq \frac{1}{N} \qty(\pdv{U}{T})_{N,\kappa} = - k_B \frac{\beta^2}{N} \qty(\pdv{U}{\beta})_{N,\kappa} $ yields, in the thermodynamic limit

\begin{equation}\label{cvlim}
    \frac{\tilde{c}_v}{k_B} =
    \left\{\begin{array}{lll}
                (\eta+2)(\eta+1) \dfrac{\Li{\eta+2}(\tilde{\xi})}{\Li{\eta+1}(\tilde{\xi})} - (\eta+1)^2 \dfrac{\Li{\eta+1}(\tilde{\xi})}{\Li{\eta}(\tilde{\xi})} & \text{for} & \beta<\beta_c \\
                (\eta+2)(\eta+1) \dfrac{\zeta(\eta+2)}{\zeta(\eta+1)} \qty(\dfrac{\beta}{\beta_c})^{-(\eta+1)}& \text{for} & \beta \geq \beta_c 
            \end{array}\right. \ .
\end{equation}
Where $\tilde{\xi}$ is the fugacity in the thermodynamic limit for $\beta<\beta_c$, which can be obtained as the solution to \eqref{LiN} with the ground state particles ignored and substituting $\beta_c$ defined in \eqref{critical}. Namely $\tilde{\xi}(\beta)$ is the solution to

\begin{equation}\label{tq}
    \Li{\eta+1}(\tilde{\xi}) = \zeta(\eta+1) \qty(\frac{\beta}{\beta_c})^{\eta+1} \ .
\end{equation}
Note that the limit $\beta \to 0$ leads to $\tilde{\xi} \to 0$ and for $\beta \to \beta_c^-$ it follows that $\tilde{\xi} \to 1$.
In order to study the non-analytical behavior of $\tilde{c}_v$, it is interesting to define the quantity

\begin{equation}\label{Deltacv}
    \frac{\Delta \tilde{c}_v}{k_B} \doteq
    \lim_{\beta \to \beta_c^+}  \frac{\tilde{c}_v}{k_B} - \lim_{\beta \to \beta_c^-}  \frac{\tilde{c}_v}{k_B} = (\eta+1)^2 \lim_{\xi\to 1^-}   \frac{\Li{\eta+1}(\xi)}{\Li{\eta}(\xi)} \ ,
\end{equation}
which is the discontinuity gap of $\tilde{c}_v$ at $\beta=\beta_c$. 
For $\eta\leq1$,  it follows that $\lim_{\xi \to 1^-}\Li{\eta}(\xi) = \infty$ and $\lim_{\xi \to 1^-}\Li{\eta+1}(\xi) = \zeta(\eta+1)$ therefore  $\Delta \tilde c_v = 0$. While for $\eta>1$, $\tilde c_v$ is discontinuous in $\beta_c$, yielding  $\Delta \frac{\tilde c_v}{k_{B}}= (\eta+1)^2 \frac{\zeta(\eta+1)}{\zeta(\eta)}$.                       

One can further study the derivative of $c_v$, using the unitless quantity
$ \frac{1}{k_B^2 \beta}\qty(\pdv{c_v}{T})_{N,\kappa} = - \frac{\beta}{k_B} \qty(\pdv{c_v}{\beta})_{N,\kappa} $
obtained from differentiating \eqref{cvlim} yielding

\begin{equation}\label{diffcvlim}
    \frac{1}{k_B^2  \beta} \qty(\pdv{\tilde{c}_v}{T})_{N,\kappa} =
    \left\{\begin{array}{lll}
                \begin{array}{ll}
                     &(\eta+2)(\eta+1)^2 \dfrac{\Li{\eta+2}(\tilde{\xi})}{\Li{\eta+1}(\tilde{\xi})} - (\eta+1)^2 \dfrac{\Li{\eta+1}(\tilde{\xi})}{\Li{\eta}(\tilde{\xi})}  \vspace{.25cm} \\
                     & \hspace{1.5cm} - (\eta+1)^3 \dfrac{(\Li{\eta+1}(\tilde{\xi}))^2 \Li{\eta-1}(\tilde{\xi})}{(\Li{\eta}(\tilde{\xi}))^3}
                \end{array}
                & \text{for} & \beta<\beta_c \vspace{.5cm}
                \\
                (\eta+2)(\eta+1)^2 \dfrac{\zeta(\eta+2)}{\zeta(\eta+1)} \qty(\dfrac{\beta}{\beta_c})^{-(\eta+1)}& \text{for} & \beta \geq \beta_c 
            \end{array}\right. \ .
\end{equation}
Similarly to \eqref{Deltacv}, one can define the discontinuity gap of the derivative of $\tilde{c}_v$ at $\beta=\beta_c$ 

\begin{equation}\label{Deltadiffcvlim}
    \frac{1}{k_B^2  \beta} \Delta \qty(\pdv{\tilde{c}_v}{T})_{N,\kappa} \doteq
    \lim_{\beta \to \beta_c^+}  \frac{1}{k_B^2  \beta}  \qty(\pdv{\tilde{c}_v}{T})_{N,\kappa}  - \lim_{\beta \to \beta_c^-} \frac{1}{k_B^2  \beta}  \qty(\pdv{\tilde{c}_v}{T})_{N,\kappa} \ .
\end{equation}
Calculating the quantity $\Delta \qty(\pdv{\tilde{c}_v}{T})_{N,\kappa}$ for $\eta \leq 0$ is inexpressive, since $\beta_c$ diverges, on the same understanding calculating $\Delta \qty(\pdv{\tilde{c}_v}{T})_{N,\kappa}$ is not representative for $\eta > 1$ since $\tilde{c}_v$ is already discontinuous. This calculation will focus on values $0<\eta\leq1$. In that regime, it follows from \eqref{diffcvlim} that the limit from the right is given by 

\begin{equation}\label{highlim}
    \lim_{\beta \to \beta_c^+}  \frac{1}{k_B^2  \beta}  \qty(\pdv{\tilde{c}_v}{T})_{N,\kappa}  = (\eta+2)(\eta+1)^2 \dfrac{\zeta(\eta+2)}{\zeta(\eta+1)}   \ .
\end{equation}
In order to calculate the equivalent limit from the left, one has to recall the series expansion in \eqref{PLseries}. Note that, for the values of $\eta$ of interest, $\Li{\eta}(\xi)$ scales as $(\log(\xi))^{\eta-1}$ as $\xi \to 1^-$, while $\Li{\eta-1}(\xi)$ scales as $(\log(\xi))^{\eta-2}$ while $\Li{\eta+1}(\xi)$ and $\Li{\eta+2}(\xi)$ converge to $\zeta(\eta+1)$ and $\zeta(\eta+2)$ respectively . Substituting the series expansion in \eqref{diffcvlim} it follows that

\begin{equation}\begin{split}\label{lowlim}
    \lim_{\beta \to \beta_c^-}  \frac{1}{k_B^2  \beta}  \qty(\pdv{\tilde{c}_v}{T})_{N,\kappa}  =\  &(\eta+2)(\eta+1)^2 \dfrac{\zeta(\eta+2)}{\zeta(\eta+1)}  \\ 
    &- (\eta+1)^3 (\zeta(\eta+1))^2\frac{\Gamma(2-\eta)}{(\Gamma(1-\eta))^3} \ \lim_{\xi\to1^-} (-\log \xi )^{1-2\eta}\ .
\end{split}\end{equation}
For $0<\eta<\half{1}$  the exponent in the last factor in \eqref{lowlim} is   positive, therefore $\Delta \qty(\pdv{\tilde{c}_v}{T})_{N,\kappa} = 0$. For $\half{1}<\eta\leq 1$  the exponent in the last factor in \eqref{lowlim} is   negative, therefore $\Delta \qty(\pdv{\tilde{c}_v}{T})_{N,\kappa} = \infty$. 
In the particular case of $\eta=\half{1}$ ---  respective to the 3-dimensional gas trapped in a box  --- the exponent vanishes, therefore

\begin{equation} \label{lowlimhalf}
     \lim_{\beta \to \beta_c^-} \frac{1}{k_B^2  \beta} \qty(\pdv{\tilde{c}_v}{T})_{N,\kappa}  = \qty(\frac{45}{8}) \frac{\Gamma(\half{5})}{\Gamma(\half{3})}- \qty(\frac{3}{2})^3 \qty(\zeta\qty(\half{3}))^2 \frac{\Gamma(\half{3})}{(\Gamma(\half{1}))^3} \approx -0.77726 \ .
\end{equation}
Leading to a convergent $\Delta \qty(\pdv{\tilde{c}_v}{T})_{N,\kappa} \approx 3.6657 k_B^2 \beta_c$ in accordance to \cite{Landau,Robertson93}. 

\begin{table}[]
\centering
\begin{tabular}{r||c|c|c|c|c|}
& $\eta \leq 0$ & $0<\eta<\half{1}$ & $ \eta = \half{1}$ & $\half{1} <\eta\leq1$ & $\eta>1$ \\ \hline  
 $\lim\limits_{\xi\to 1^-}   {\Li{\eta+1}(\xi)}$ & {Divergent ($+\infty)$ } & \multicolumn{4}{c|}{Convergent $\zeta(\eta+1)$ }  \\  \hline
 $\lim\limits_{\xi\to 1^-}   {\Li{\eta}(\xi)}$& \multicolumn{4}{c|}{Divergent ($+\infty)$ } & {Convergent $\zeta(\eta)$ }                        \\ \hline
\tableheight $\beta_c$& {Divergent ($+\infty)$ } & \multicolumn{4}{c|}{Convergent --- see \eqref{critical} }                                         \\  \hline
\tableheight $\tilde{c}_v$&       ---        &  \multicolumn{3}{c|}{Continuous at $\beta_c$ }                        &Discontinuous at $\beta_c$   \\ \hline
 \tableheight $\Delta \tilde{c}_v$&       ---        &  \multicolumn{3}{c|}{$0 $ }                        & $k_B (\eta+1)^2 \frac{\zeta(\eta+1)}{\zeta(\eta)} $  \\ \hline 
 $\qty(\pdv{\tilde{c}_v}{\beta})_{N,\kappa}$ &      ---         &{Continuous at $\beta_c$}& \multicolumn{2}{c|}{Discontinuous at $\beta_c$}&   ---       \\ \hline
  $\Delta \qty(\pdv{\tilde{c}_v}{\beta})_{N,\kappa}$ &      ---         & 0 & $\approx 3.66 \ k_B^2 \beta_c$ &$+\infty$&   ---       \\ 
\end{tabular}
\caption{Diagram relating the convergence of polylogarithms in terms of the density of states exponent $\eta$ to the convergence of the critical temperature, $\beta_c$ in \eqref{critical}. It is also presented how, as a consequence, the value of $\eta$ affects the continuity of specific heat, $\tilde{c}_v$ in \eqref{cvlim}, and of its derivative in \eqref{Deltadiffcvlim}.}
    \label{summary}
\end{table}

A summary for the convergence of  $\beta_c$ along with the discontinuities of $\tilde{c}_v$ and $\qty(\pdv{\tilde{c}_v}{T})_{N,\kappa}$, in terms of $\eta$ are presented in Table \ref{summary}. 
With the study of the thermodynamic limit for general values of $\eta$ done, the following section studies BE statistics for a finite number of particles.

\section{BE statistics for a finite number of particles}\label{finito}
\paragraph{}
In order to appropriately study BE condensation in terms of $N$, one needs to express the thermodynamical quantities of interest --- namely the fraction of particles in the ground state, specific heat and its derivative --- in a manner that is appropriate to compare to the critical temperature.
Since $\beta_c$ in \eqref{critical} is defined in terms of $N$, one has to write $n_0$ and $U$ in terms of $\beta$ and $N$, since these were written in \eqref{Averages4} in terms of $\beta$ and $\xi$ it would suffice to obtain $\xi$ as a function of $\beta$ and $N$.  From \eqref{LiN} it is straightforward to see that $N$ is strictly increasing with $\xi \in [0,1)$ therefore $\xi(\beta,N)$ is well defined  as the inverse of \eqref{LiN}. 

To the best of my knowledge, $\xi(\beta,N)$ has never been written in closed analytical form. However as the inverse of a strictly increasing function $\xi(\beta,N)$ can be implemented through simple numerical algorithms. An implementation of it can be seen in the IGQG python library, available in my GitHub repository \cite{github}. This implementation is based on the mpmath library \cite{mpmath} that allows for calculations of arbitrary precision. All thermodynamical quantities of interest can be exactly written in terms of $\xi(\beta,N)$ --- which will be written only as $\xi$ in this section for simpler notation. These calculations have not been published before to the best of my knowledge. Graphs for the thermodynamical quantities obtained from this implementation will be presented below. It is observed that for finite N the non analytical behaviour disappears, which is in accordance with the fact that $U$ and $N$ in \eqref{Averages4} are continuous for positive $\beta$ and $0\leq \xi <1$.

\begin{figure}
    \centering
    \includegraphics[width=.8\textwidth]{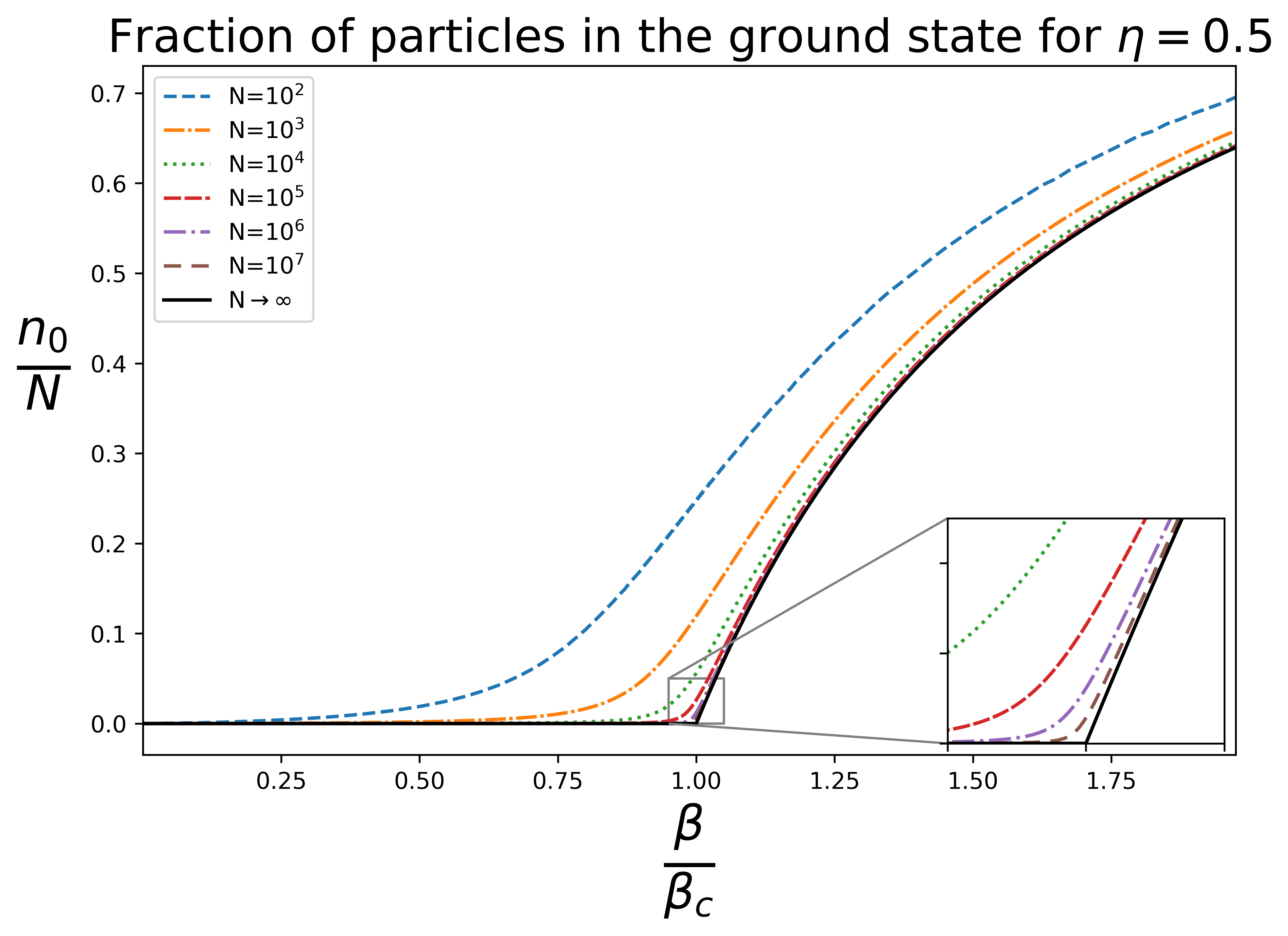}    \includegraphics[width=.8\textwidth]{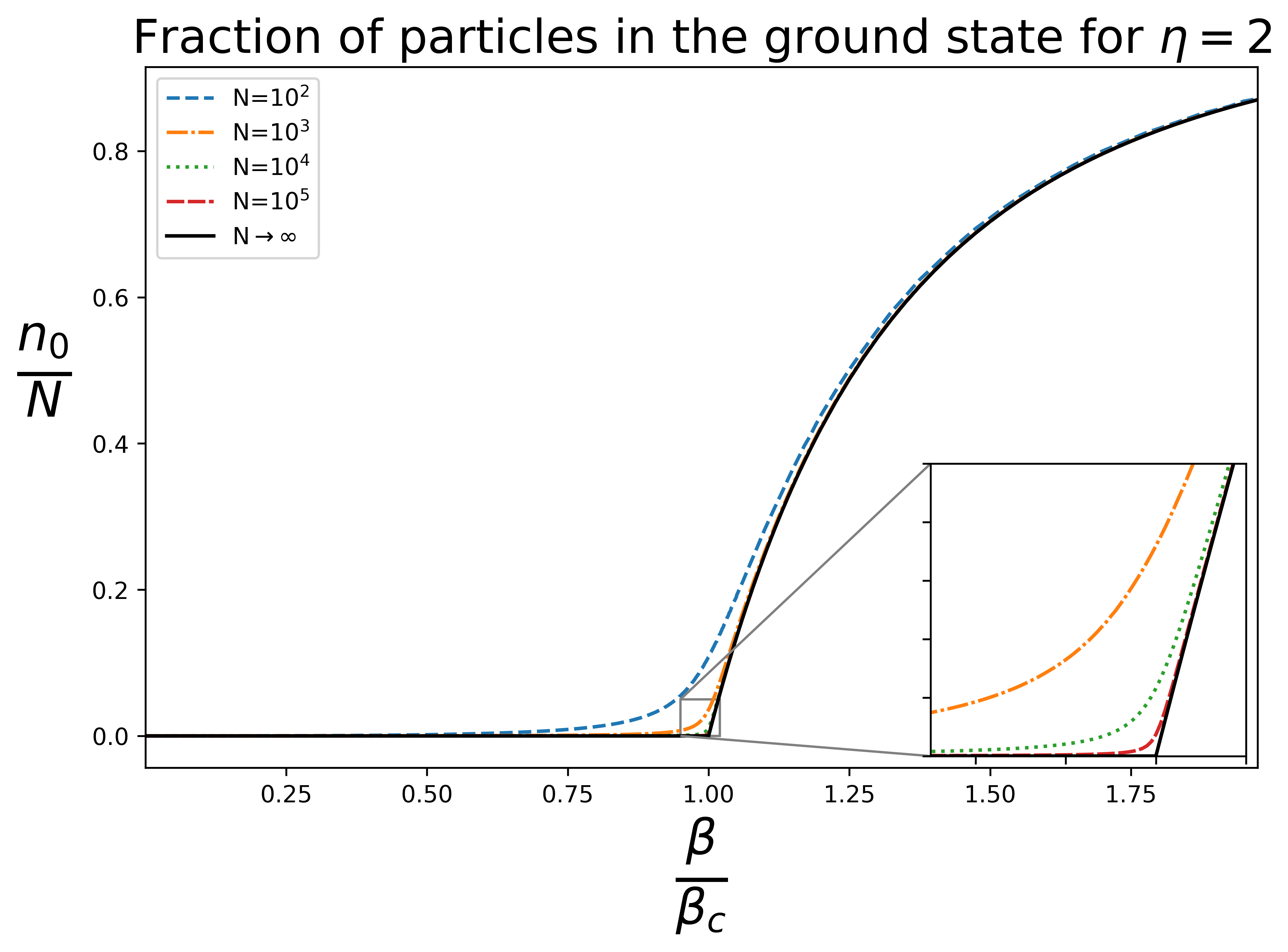}
    \caption{Fraction of particles in the ground state \eqref{finalfraction} for $\eta = \half{1}$ and $\eta = 2$. On both cases it can be seen how the fraction of particle in the ground state, for finite $N$, approach, smoothly, the non-analytical curve for the thermodynamic limit. }
    \label{fig:fraction}
\end{figure}

The fraction of particles in the ground state can be obtained by dividing \eqref{LiN} by the number of particles $N$ and substituting $\beta_c$ as in \eqref{critical} obtaining

\begin{equation}\label{finalfraction}
    \frac{n_0}{N} = 1- \frac{\Li{\eta+1}(\xi)}{\zeta(\eta+1)} \qty(\frac{\beta}{\beta_c})^{-(\eta+1)} \ .
\end{equation}
Above it is supposed a value of $\eta$ for which $\beta_c$ converges, hence  $\lim\limits_{\xi \to 1^-}\Li{\eta+1}(\xi)$ is taken to be $\zeta(\eta+1)$. Graphs for $\frac{n_0}{N} $ are presented for the 3-dimensional gas in a box ($\eta = \half{1}$) and the 3-dimensional harmonically trapped gas ($\eta = 2$) in Fig. \ref{fig:fraction} with a comparison to the fraction of particles calculated in the thermodynamic limit \eqref{cvlim}.

\begin{figure}
    \centering
    \includegraphics[width=.8\textwidth]{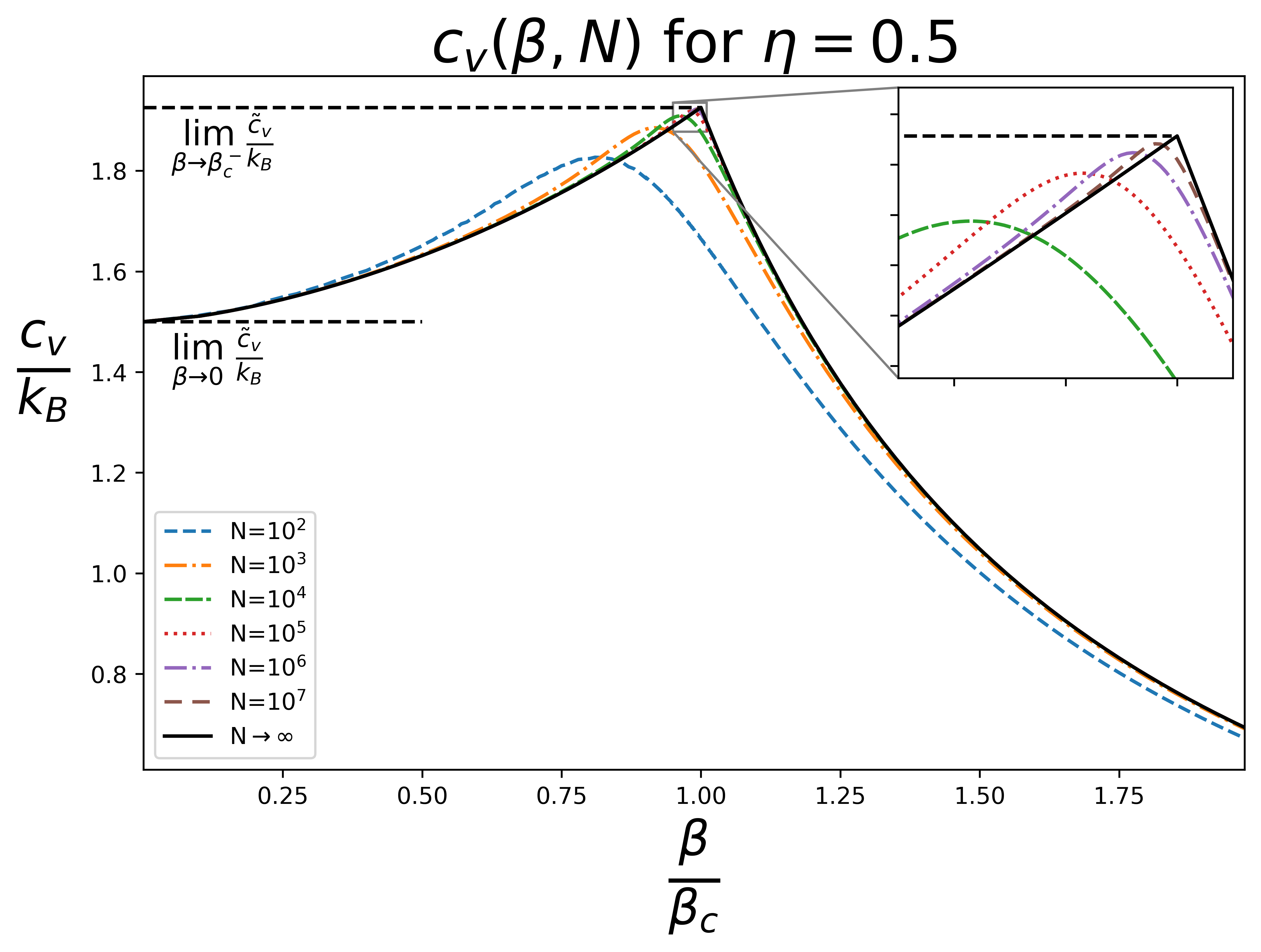}    \includegraphics[width=.8\textwidth]{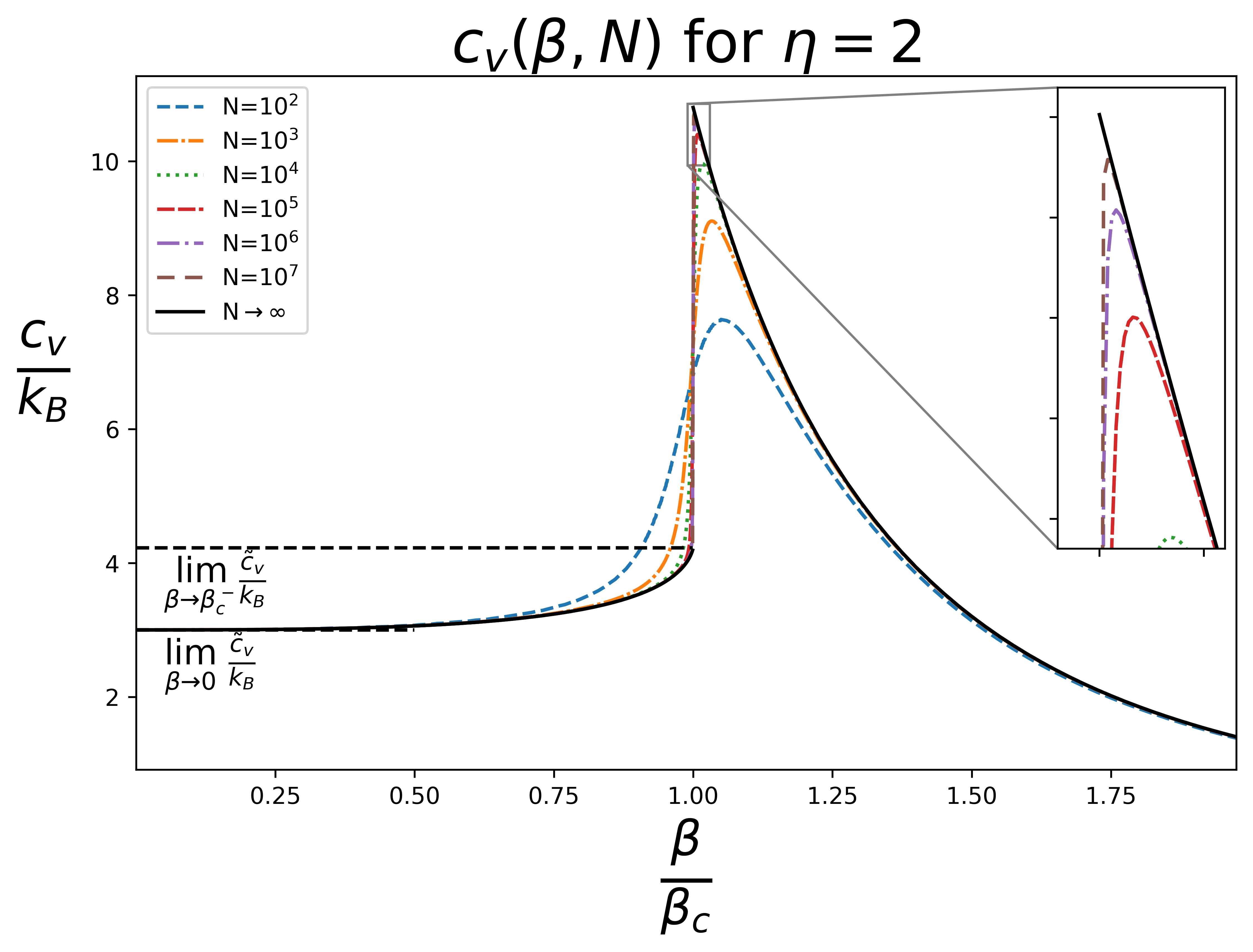}
    \caption{Specific heat of a Bose gas \eqref{cv} for $\eta = \half{1}$ and $\eta = 2$. In the first picture is seen that the quantity is continuous, albeit not smooth, as $\beta$ approaches $\beta_c$ in the thermodynamic limit. On the second one is seen how in the thermodynamic limit, the value diverges is discontinuous when $\beta$ approaches $\beta_c$ from the left. On both cases it can be seen how the non-analytical curve is approached from the curves of specific heat for finite $N$.}
    \label{fig:cv}
\end{figure}

The specific heat, $c_v \doteq \frac{1}{N} \qty(\pdv{U}{T})_{N,\kappa} = - k_B \frac{\beta^2}{N} \qty(\pdv{U}{\beta})_{N,\kappa} $, can be calculated from the direct differentiation of $U$ in \eqref{LiU}, obtaining

\begin{equation}\label{cv}
    \frac{c_v}{k_B} =  \frac{\kappa}{N} \frac{\Gamma(\eta+2)}{\beta^{\eta+1}}\qty[ (\eta+2) \ \Li{\eta+2}(\xi) - \frac{\beta}{\xi} \qty(\pdv{\xi}{\beta})_{N,\kappa} \Li{\eta+1}(\xi)  ] \ ,
\end{equation}
where $\qty(\pdv{\xi}{\beta})_{N,\kappa}$ can be obtained from the implicit differentiation of $N$ in \eqref{LiN} with respect to $\beta$, yielding

\begin{equation}\label{dxidb}
    \frac{1}{\xi}\qty(\pdv{\xi}{\beta})_{N,\kappa} =
    \frac{ \kappa
    \frac{ \Gamma(\eta+2)}{\beta^{\eta+2}} \Li{\eta+1}(\xi) }{\kappa \frac{\Gamma(\eta+1)}{\beta^{\eta+1}} \  \Li{\eta}(\xi)  + \frac{\xi}{(1-\xi)^2}} \ .
\end{equation}
Note that the ground state contribution to $c_v$ appears in the second term of the denominator of \eqref{dxidb}. Graphs for $c_v$ are presented for $\eta = \half{1}$ and $\eta=2$ in Fig. \ref{fig:cv}. 
The comparison to the thermodynamic limit is obtained from \eqref{cvlim}. For $\beta \geq \beta_c$ it follows that $\tilde{c}_v$ can be calculated directly in terms of $\beta$ directly, for $\beta<\beta_c$ the graphed values of $\tilde{c}_v$ are based on the numerical implementation of the solution of \eqref{tq} also found in \cite{github}.
It is particularly interesting to notice, in  Fig. \ref{fig:cv}, how the  discontinuity in $\tilde{c}_v$ for $\eta=2$ is approached from the continuous values calculated for finite $N$.

\begin{figure}
    \centering
    \includegraphics[width=.85\textwidth]{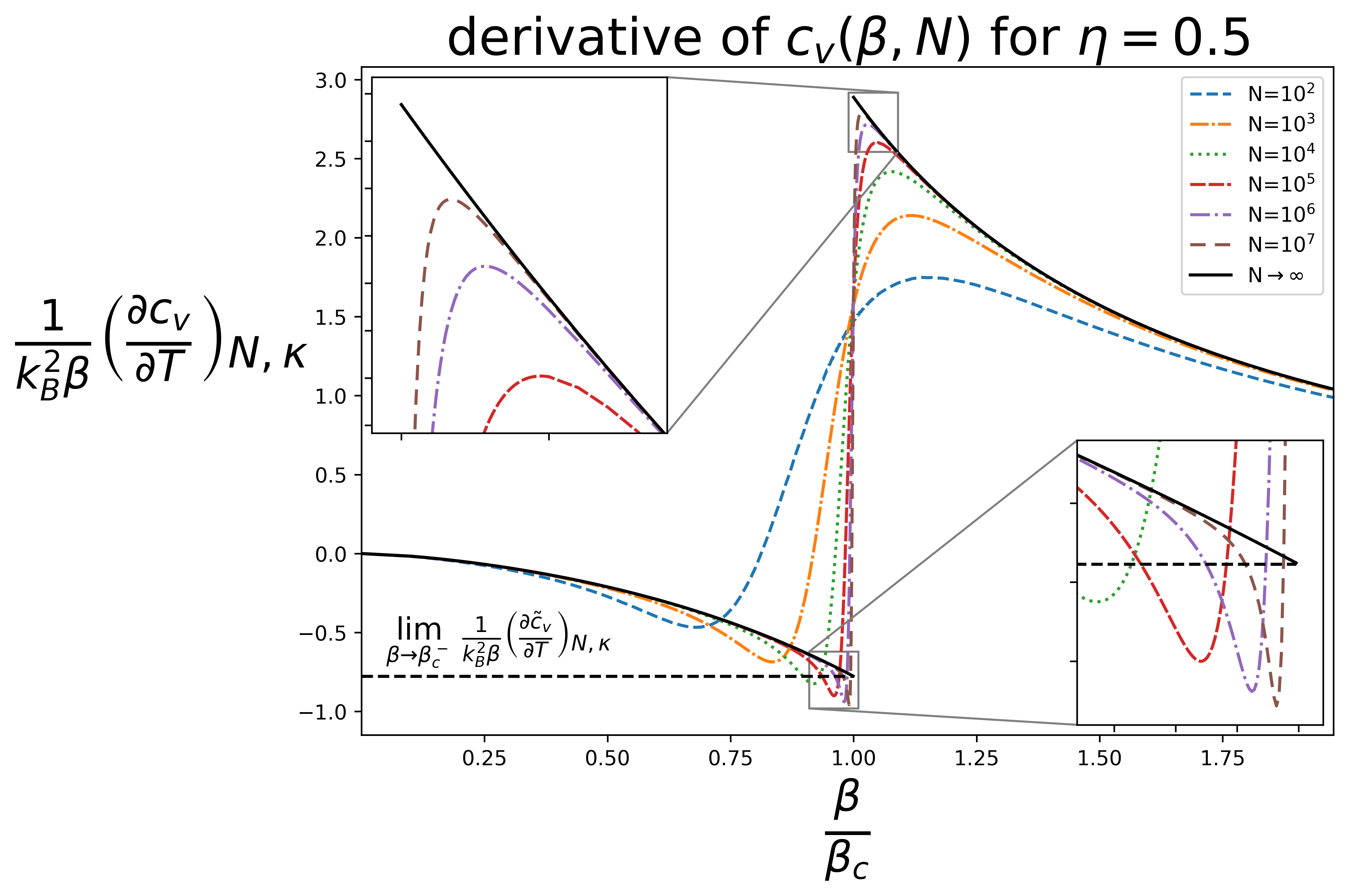}    \includegraphics[width=.85\textwidth]{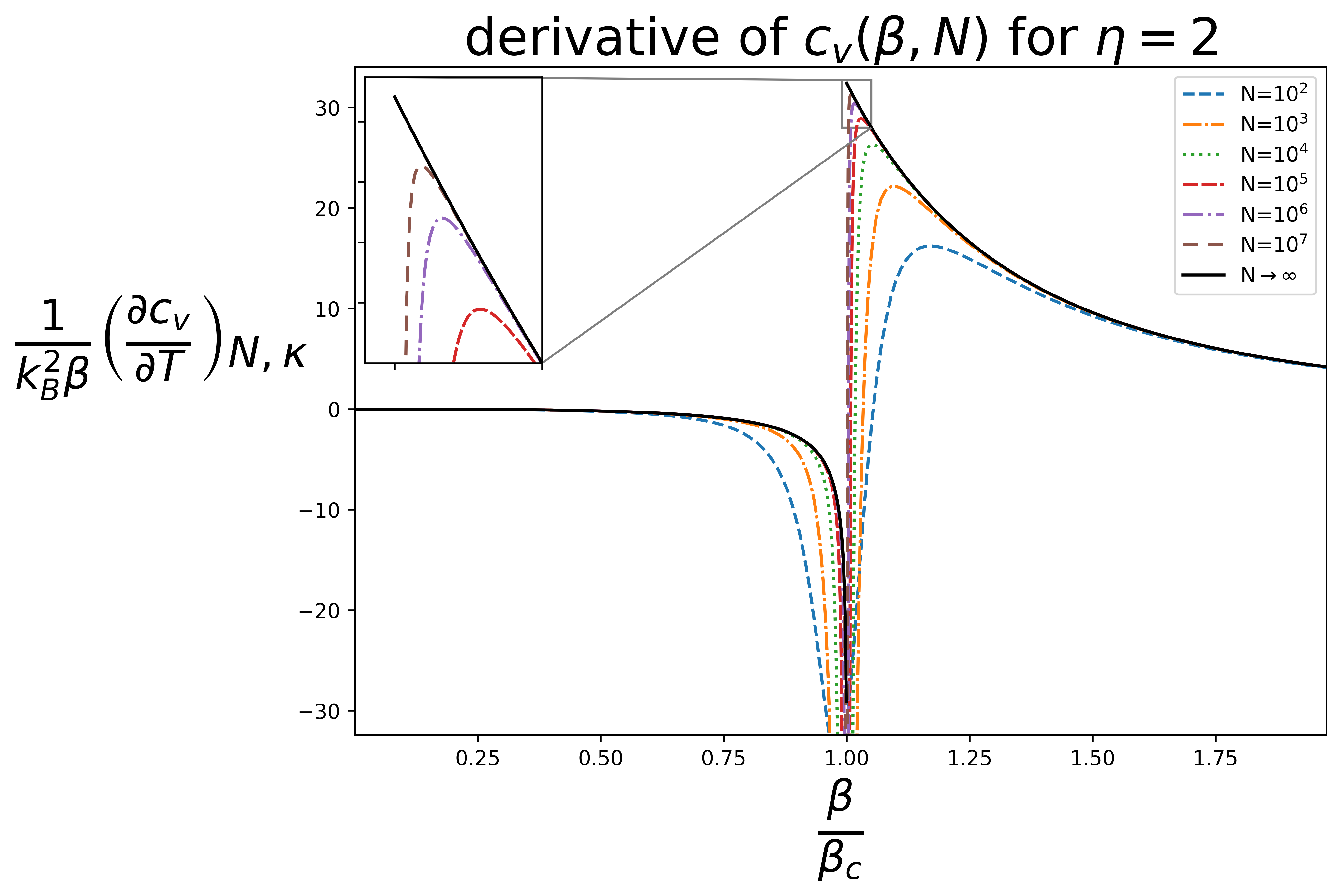}
    \caption{Graphs for $\qty(\pdv{c_v}{T})_{N,\kappa}$ of a Bose gas \eqref{dcvdtob} for $\eta = \half{1}$ and $\eta = 2$. On the first image it is seen that the quantity becomes smaller than the one expected from calculations in the thermodynamic limit. On the second one is seen how in the thermodynamic limit, the value diverges to $-\infty$ when $\beta$ approaches $\beta_c$ from the left.}
    \label{fig:dcv}
\end{figure}

To further compare the study of BE statistics for finite $N$ to the one in the thermodynamic limit, it is important to study the derivative of $c_v$. Again, this can be done by the study of the unitless quantity
$ \frac{1}{k_B^2 \beta}\qty(\pdv{c_v}{T})_{N,\kappa} = - \frac{\beta}{k_B} \qty(\pdv{c_v}{\beta})_{N,\kappa} $ which can be
obtained from differentiating \eqref{cv} yielding

\begin{equation}\begin{split} \label{dcvdtob}
    \frac{1}{k_B^2  \beta} \qty(\pdv{c_v}{T})_{N,\kappa} = &\frac{\kappa}{N}   \frac{\Gamma(\eta+2)}{\beta^{\eta+1}}
    \left[ (\eta+2) (\eta+1) \ \Li{\eta+2}(\xi)   - \frac{\beta}{\xi}\qty(\pdv{\xi}{\beta})_{N,\kappa}  2 (\eta+1) \ \Li{\eta+1}(\xi) \right. \\
    & \ \left. + \qty(\frac{\beta}{\xi}\qty(\pdv{\xi}{\beta})_{N,\kappa})^2  \ \qty( \Li{\eta}(\xi) - \Li{\eta+1}(\xi) )  +  \frac{\beta^2}{\xi}\qty(\pdv[2]{\xi}{\beta})_{N,\kappa}  \ \Li{\eta+1}(\xi)     \right] \ ,
\end{split}\end{equation}
where $\qty(\pdv{\xi}{\beta})_{N,\kappa}$ was already calculated in \eqref{dxidb} and $\qty(\pdv[2]{\xi}{\beta})_{N,\kappa}$ can similarly be obtained from the second implicit differentiation of $N$ in \eqref{LiN} with respect to $\beta$, yielding

\begin{equation}\begin{split}\label{dxi2db2}
    \frac{1}{\xi} \qty(\pdv[2]{\xi}{\beta})_{N,\kappa} =
    -& \left[ 
     \kappa\frac{\Gamma(\eta+3)}{\beta^{\eta+3}} \Li{\eta+1}(\xi) 
    - \frac{1}{\xi}\qty(\pdv{\xi}{\beta})_{N,\kappa} 2  \kappa\frac{\Gamma(\eta+2)}{\beta^{\eta+2}} \Li{\eta}(\xi) \right. \\
    & \ \ + \qty(\frac{1}{\xi}\qty(\pdv{\xi}{\beta})_{N,\kappa})^2 \   \kappa\frac{\Gamma(\eta+1)}{\beta^{\eta+1}} \ \qty( \Li{\eta-1}(\xi) - \Li{\eta}(\xi) ) \\
    & \ \  \left.  + 2  \qty(\frac{1}{\xi}\qty(\pdv{\xi}{\beta})_{N,\kappa})^2 \frac{\xi^2}{\qty(1-\xi)^3} \right] \times \left[{\kappa \frac{\Gamma(\eta+1)}{\beta^{\eta+1}} \  \Li{\eta}(\xi)  + \frac{\xi}{(1-\xi)^2}} \right]^{-1} \ .
\end{split}\end{equation}
Graphs for the unitless quantity
$ \frac{1}{k_B^2 \beta}\qty(\pdv{c_v}{T})_{N,\kappa}$ are presented for $\eta = \half{1}$ and $\eta=2$ in Fig. \ref{fig:dcv}.

An interesting non intuitive behaviour  becomes clear in  Fig. \ref{fig:dcv}.  For $\eta =\half{1}$, it can be seen that the value of  $\frac{1}{k_B^2  \beta} \qty(\pdv{c_v}{T})_{N,\kappa}$ grows smaller than the  minimum possible value obtained in the thermodynamic limit, given as $\lim\limits_{\beta \to \beta_c^-} \frac{1}{k_B^2  \beta} \qty(\pdv{\tilde{c}_v}{T})_{N,\kappa}$ in \eqref{lowlimhalf}.  Such behavior is not observed for $\eta <\half{1}$ --- since at those values $\qty(\pdv{\tilde{c}_v}{T})_{N,\kappa}$ is continuous ---  it is also not observed for $\eta>\half{1}$ --- since, per \eqref{lowlim}, it follows that $\lim\limits_{\beta \to \beta_c^-} \frac{1}{k_B^2  \beta} \qty(\pdv{\tilde{c}_v}{T})_{N,\kappa}$ goes to negative infinite.

This implies that the thermodynamic limit, as presented in Sec. \ref{limitsection}, misses interesting physical behaviour.  
Mainly, the calculation of $\tilde{c}_v$ was made assuming that there are no particles in the ground state for $\beta < \beta_c$ -- as is also the case for all calculations made in Sec. \ref{limitsection}. The fact that $\qty(\pdv{\tilde{c}_v}{T})_{N,\kappa}$ obtained this way is strictly decreasing for $\beta<\beta_c$ indicated that the discontinuity at $\beta_c$ is approached from above. 
The behaviour found in Fig. \ref{fig:dcv} indicates that accounting for $n_0$ above the critical temperature leads to a smaller value of  $\qty(\pdv{{c}_v}{T})_{N,\kappa}$.
Therefore the discontinuity found in the thermodynamic limit is approached from below, not above. 

\begin{figure}[h]
\centering
    \includegraphics[width=.65\textwidth]{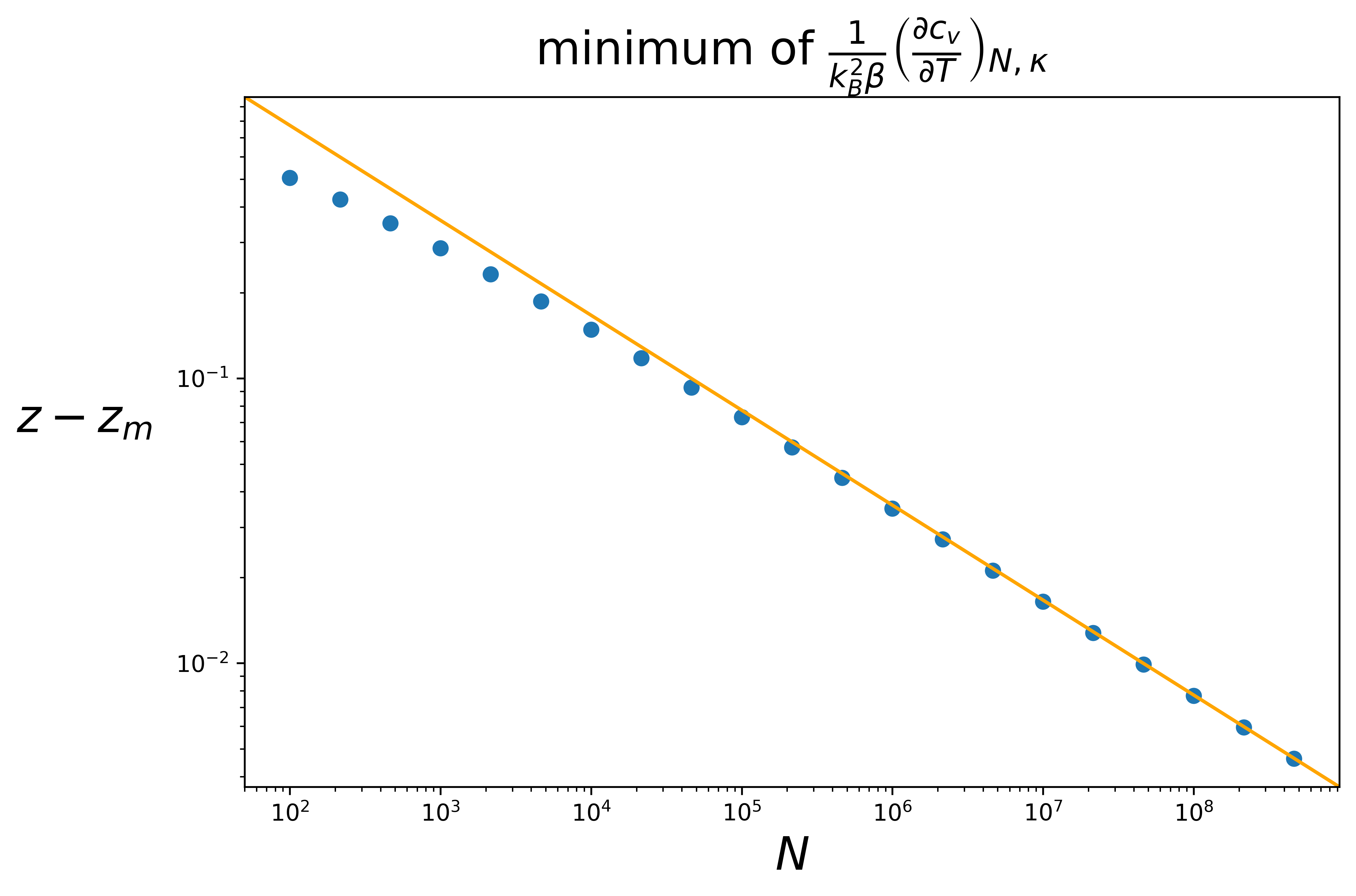}
    \caption{Graph for the value of $z$, defined in \eqref{zed}, calculated numerically for $N$ ranging from $10^2$ to $10^8$ (scattered blue points) and the approximation in order of $N^{-\third{1}}$  (solid orange line)  ---  meaning $z(N) = z_{m} + \bar{z} N^{-\third{1}}$ as in \eqref{zed}.  }
    \label{fig:zed}
\end{figure}

Interestingly, this result is supported by analytical calculations. For $\eta=\half{1}$, the minimal value of $\qty(\pdv{\tilde{c}_v}{T})_{N,\kappa}$ --- calculated without assuming $n_0=0$ below $\beta_c$ --- is related to $N$ as 

\begin{equation} \label{zed}
    z(N) \doteq \min_\beta \qty(\pdv{\tilde{c}_v}{T})_{N,\kappa} = z_{m} + \bar{z} N^{-\third{1}} + o(N^{-\third{1}}) \ ,
\end{equation}
where 
\begin{equation} \label{zedvalues}
    z_m \approx -0.97337  \qq{and} \bar{z} \approx 3.5881 \ ,
\end{equation}
and where $o$ stands for the smaller order notation, $\lim\limits_{N\to\infty} \frac{o \qty(N^{-\third{1}})}{N^{-\third{1}}} = 0$.
Note that, as expected from Fig. \ref{fig:dcv}, $\lim\limits_{N\to\infty} z(N) = z_m < \lim\limits_{\beta \to \beta_c^-} \frac{1}{k_B^2  \beta} \qty(\pdv{\tilde{c}_v}{T})_{N,\kappa} $ calculated in  \eqref{lowlimhalf}. 
The analytical calculation proving \eqref{zed} is presented in Appendix \ref{appendixdcv}. A comparison between the values of $Z(N)$ calculated numerically compared to the ones given by \eqref{zed} in the order of $N^{-\third{1}}$ is presented in Fig. \ref{fig:zed}.

\begin{figure}
    \includegraphics[width=.75\textwidth]{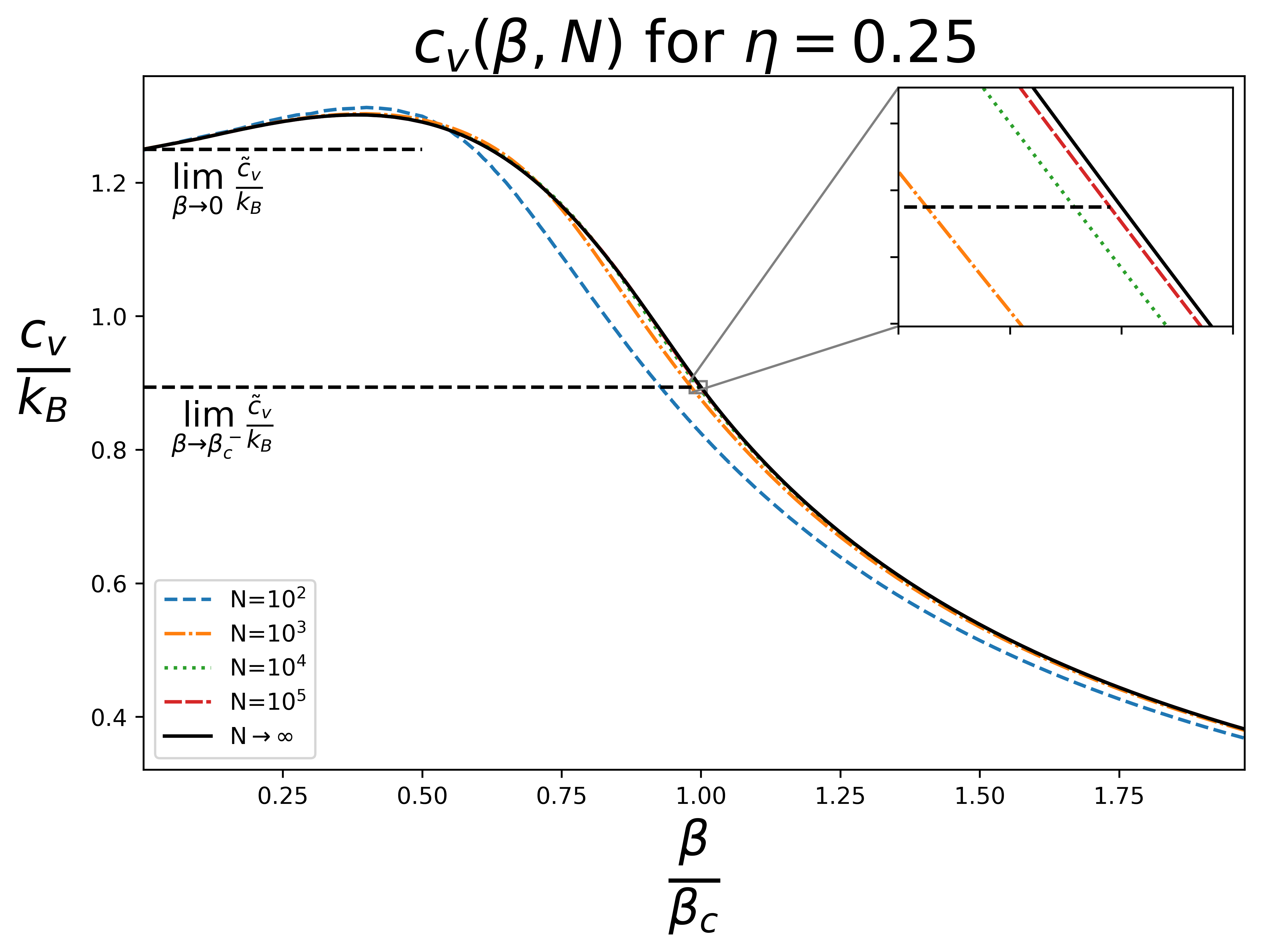} 
    \centering
    \includegraphics[width=.85\textwidth]{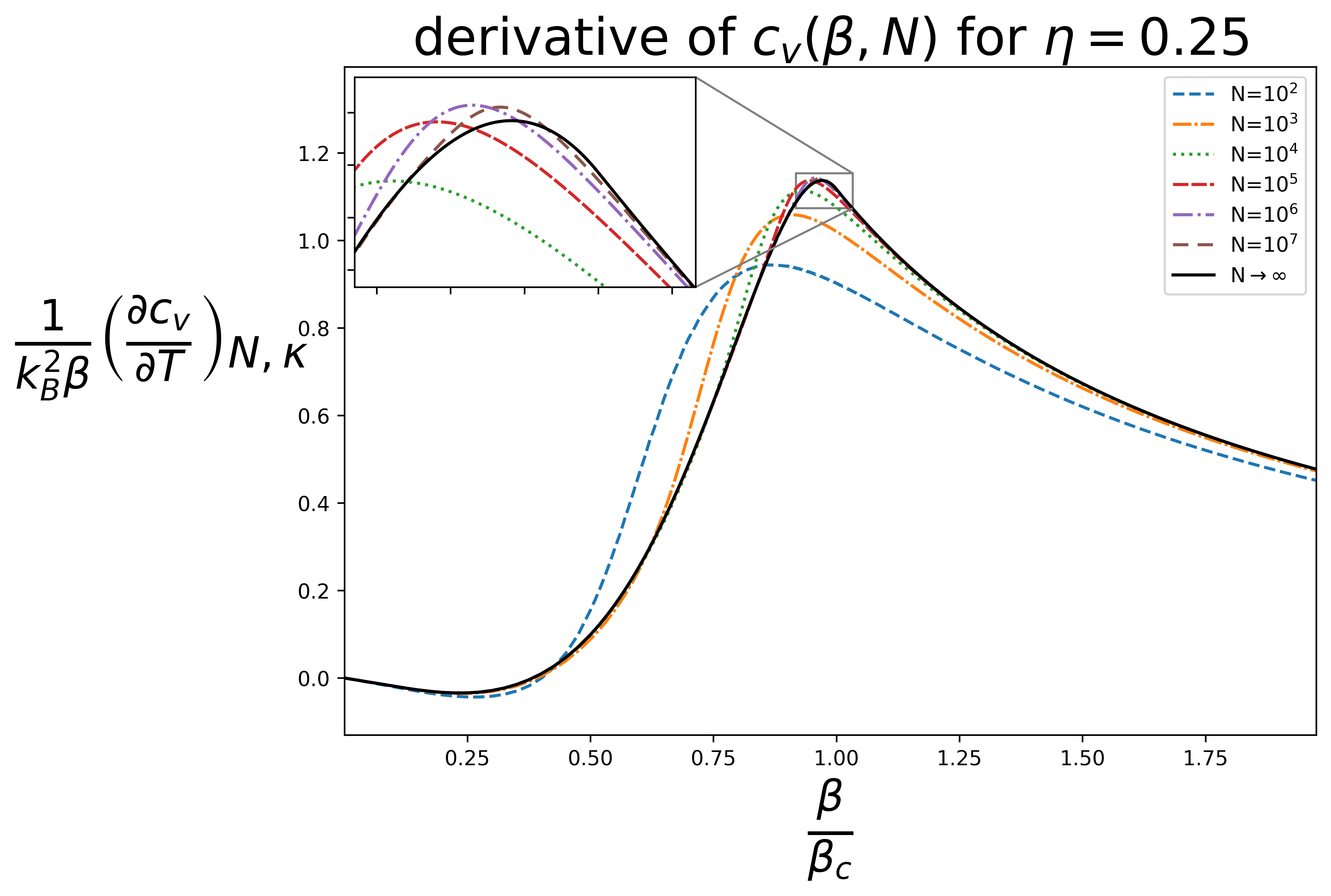}
    \caption{Graphs for $c_v$ in \eqref{cv} and $\qty(\pdv{c_v}{T})_{N,\kappa}$ in \eqref{dcvdtob} of a Bose gas  for $\eta = \nfrac{1}{4}$. Unlike in the previous pictures, we see that both quantities are smooth for this value of $\eta$. }
    \label{fig:eta.25}
\end{figure}

Other interesting properties can be observed from the study of Bose gases in finite N. As commented in Sec. \ref{limitsection}, $c_v$ and its derivative are continuous for $0<\eta<\half{1}$. The same numerical investigation used in Figs. \ref{fig:fraction} - \ref{fig:dcv} can also illustrate an important difference in qualitative behaviour for this regime.
In Fig. \ref{fig:eta.25} the graphs for $c_v$ and $\qty(\pdv{c_v}{T})_{N,\kappa}$ are presented for $\eta = \nfrac{1}{4}$ --- which is equivalent, per \eqref{ketapl}, to a quartic interaction ($\tau=4$) in a 2-dimensional gas.
It is interesting to see that for $\eta = \nfrac{1}{4}$, the specific heat at $\beta =0$ is larger than for $\beta \to \beta_c^-$. In this case $c_v$ is increasing for small $\beta$ --- in accordance to \eqref{cvlim} --- but as $\beta$ increases, it reduces smoothly --- as expected from Table \ref{summary} --- so no non-analytical behaviour is observed for $c_v$ or its first  derivative at $\beta_c$.

\section{Conclusions}
\paragraph{}
The present article presents a complete description of BE statistics that does not rely on the thermodynamic limit. This is made possible from the numerical calculation of $\xi(\beta,N)$ as the inverse of \eqref{LiN}. From this, all thermodynamical quantities can be written in terms of $\beta$ and $N$ allowing for a direct comparison to $\beta_c$.

The thermodynamical quantities that identify the BE condensation were calculated here. The fraction of particles in the ground state is calculated exactly, for arbitrary $\eta$, in terms of $\xi(\beta,N)$ in \eqref{finalfraction}. Supplemented by the numerical inversion,  graphs of this quantity for a gas trapped in a regular box ($\eta=\half{1}$) and in a harmonic potential ($\eta=2$) are presented in Fig. \ref{fig:fraction}. Similarly the specific heat is calculated in \eqref{cv} and the numerical results are presented in Fig. \ref{fig:cv}. Finally, the derivative of specific heat is calculated in \eqref{dcvdtob} and presented in Fig. \ref{fig:dcv}. 

In all of these figures, the thermodynamical quantities were calculated for values of $N$ raging from $10^2$ to $10^7$ -- in accordance to the numbers found in the experimental observation of BE condensation -- where significant differences are observed in comparison to the calculations made in the thermodynamic limit. A summary for the convergence and continuity of these quantities in the thermodynamic limit is presented at Table \ref{summary}.
{These graphs by themselves an important visualization of the role of the thermodynamic limit for the non analytical behaviour indicating the phase transition in BE gases, hence an important pedagogical tool for the study of phase transitions.}

Particularly in Fig. \ref{fig:dcv}, a fundamental difference in the qualitative behaviour of specific heat derivative is observed. Considering particles in the ground state below critical temperature, the minimum value of this quantity is smaller than the one given in the thermodynamic limit. 
This result is supported by analytical calculations \eqref{zed} and it is found --- both by numerical results in Fig. \ref{fig:zed} and analytical calculations in \eqref{zed} --- that such minimum value scales with $N^{-\third{1}}$.

With the numerical inversion of \eqref{LiN} established and available at \cite{github}, further studies on BE condensation for a broad range of quantum systems --- whenever the density of states exponent $\eta$ can be identified --- are now possible without relying on the thermodynamic limit nor on specific approximations --- the method presented here obtains calculations with arbitrary precision for any value of $\beta$ and $N$. 
Future work may entail a precise calculation for other definitions of BE critical temperature \cite{Jaouadi11,Noronha13} and the information geometry of quantum gases\cite{Pessoa21,LopezPicon21}.

\section*{Acknowledgments}\paragraph{}
I would like to acknowledge the much appreciated assistance of D. Robbins, with whom the calculations in Appendix \ref{appendixdcv} were performed. D. Robbins was also the first person to introduce me to the polylogarithm series expansion presented in \eqref{PLseries}. I would also like to thank B. Arderucio Costa, A. Caticha, C. Cafaro, and R. Correa da Silva for important discussions during the development of the present article. 

\section*{Appendix}
\appendix

\section{Maximum entropy in Fock spaces}\label{appendixMaxEnt}
\paragraph{}
This appendix  will derive the grand canonical Gibbs distribution \eqref{BEGibbs} for BE statistics and obtain the expected values for the number of particles and internal energy found in \eqref{Averages}. 
Probabilities in statistical mechanics \cite{Jaynes57,Jaynes65} are assigned by finding the probability $\rho(x)$ that maximizes entropy, $\rho(x) = \arg\max\limits_p S[p|q]$ {where} 

\begin{equation}\label{entropy}
    S[p|q] = - \sum_x p(x) \log\left(\frac{p(x)}{q(x)}\right) \ ,
\end{equation}
with constraints on the form of expected values of the sufficient statistics, $a_\alpha$,

\begin{equation}\label{values}
    A_\alpha = \sum_x p(x) a_\alpha(x) \ ,
\end{equation}
and normalization. This maximization is achieved by the Gibbs distribution

\begin{equation}\label{Gibbsdefinition}
    \rho(x|\lambda) = \frac{q(x)}{Z(\lambda)} \exp \left(-\sum_{\alpha} \lambda_\alpha a_\alpha(x)\right) \ ,
\end{equation}
where each $\lambda_\alpha$ is the Lagrange multiplier related to the  expected value constraints in $a_\alpha$. $Z(\lambda)$ is the partition function, a normalization factor independent of $x$. The expected values in \eqref{values} can be recovered as $A_\alpha = - \pdv{}{\lambda_\alpha} \log Z$.

In Fock spaces,  $x=\{x_i\}$, the measure is given as $\sum\limits_x=\sum\limits_{x_1 =0}^\infty \ \sum\limits_{x_2 =0}^\infty \ldots \ $, and $q(x)$ in uniform. In the grand canonical ensemble the sufficient statistics are chosen as the energy $a_1(x) = \sum_i \epsilon_i x_i $ and the total number of particles $a_2(x) = \sum_i  x_i $. The Gibbs distribution is, therefore, of the form

\begin{equation}
    \rho(x|\lambda_1,\lambda_2) = \frac{1}{Z(\lambda)} \prod_i e^{-\lambda_1  \epsilon_i x_i} \ e^{ -\lambda_2 x_i} \ ,
\end{equation}
where $\lambda_1$ is identified as $\beta$ and $\lambda_2$ is identified as $-\beta\mu$ or, equivalently, $\xi = e^{-\lambda_2}$, leading to  \eqref{BEGibbs}. The normalization factor is identified as

\begin{equation}
    Z(\beta,\xi) = \sum_x \prod_i e^{-\beta  \epsilon_i x_i} \ \xi^{x_i} = \prod_i \left( 1 -  e^{-\beta \epsilon_i } \xi \right)^{-1} \ ,
\end{equation}
Leading to the expected values \eqref{values}

\begin{subequations} \label{values2}
\begin{align}
    A_1 = U & =\sum_ i  {\xi \  \epsilon_i }  \left( e^{\beta \epsilon_i } - \xi \right)^{-1}  \label{Uap}   \\
    A_2 = N & =\sum_ i  { \xi } \left( e^{\beta \epsilon_i } -   \xi \right)^{-1}  \label{Nap} \ ,
\end{align} \end{subequations}
which are equivalent to \eqref{sumU} and \eqref{sumN} respectively.

\section{Calculations in the thermodynamic limit} \label{appendixcv}
\paragraph{}
This appendix will derive the thermodynamical quantities of interest -- calculated for finite $N$ in \eqref{finalfraction}, \eqref{cv} and \eqref{dcvdtob} -- reduce to the ones presented in the thermodynamic limit -- respectively \eqref{tildefraction}, \eqref{cvlim} and \eqref{diffcvlim}. 

As explained in Sec. \ref{limitsection} $\beta<\beta_c$ it implies $n_0 = 0$ in the thermodynamic limit. It follows directly that $\frac{n_0}{N}=0$, reducing to \eqref{tildefraction} for $\beta<\beta_c$. From implicit differentiation of \eqref{LiN}, it follows that

\begin{equation}\label{dxidbTL}
    \frac{1}{\xi}\qty(\pdv{\xi}{\beta})_{N,\kappa} =
    \frac{1}{\beta} \ (\eta+1) \frac{\Li{\eta+1}(\xi)}{\Li{\eta}(\xi)} \ ,
\end{equation}
which is equivalent to \eqref{dxidb} in the thermodynamic limit for $\beta<\beta_c$. 
Similarly, it follows from the second differentiation of \eqref{LiN} that

\begin{equation}\begin{split}\label{dxi2db2TL}
    \frac{1}{\xi}\qty(\pdv[2]{\xi}{\beta})_{N,\kappa} =
    - \frac{1}{\beta^2} & \left[  (\eta+2)(\eta+1)\frac{\Li{\eta+1}(\xi)}{\Li{\eta}(\xi)} -(\eta+1)^2 \frac{\Li{\eta+1}(\xi)}{\Li{\eta}(\xi)} \right. \\
    &\left. \ \ -  (\eta+1)^2 \ \frac{\qty(\Li{\eta+1}(\xi))^2}{\qty(\Li{\eta}(\xi))^3}  \ \qty( \Li{\eta-1}(\xi) - \Li{\eta}(\xi) ) \right] \ ,
\end{split}\end{equation}
equivalent to \eqref{dxi2db2} in the thermodynamic limit for $\beta<\beta_c$. Substituting \eqref{dxidbTL} into \eqref{cv} one obtains \eqref{cvlim} for $\beta<\beta_c$ and substituting  \eqref{dxidbTL} and \eqref{dxi2db2TL} into \eqref{dcvdtob} one obtains \eqref{diffcvlim} for $\beta<\beta_c$,  completing the calculation.

As $\beta \geq \beta_c$ it implies $\xi=1$ in the thermodynamic limit. 
Substituting $\beta_c$ in \eqref{critical} into \eqref{finalfraction} becomes \eqref{tildefraction} for $\beta \geq \beta_c$.
Analogously, substituting $\xi=1$ into \eqref{dxidb} and \eqref{dxi2db2}  it follows directly that $\qty(\pdv{\xi}{\beta})_{N,\kappa} = 0$. Therefore, \eqref{cv} becomes \eqref{cvlim} for $\beta \geq \beta_c$ and from the direct differentiation of  \eqref{cvlim} one obtains \eqref{diffcvlim} for $\beta \geq \beta_c$, completing the calculation.

\section{Minimum value of  $\qty(\pdv{c_v}{T})_{N,\kappa}$ }\label{appendixdcv}
\paragraph{}
This appendix will derive \eqref{zed} analytically by an expansion of \eqref{dcvdtob}, thus explaining the observance of values of $\qty(\pdv{c_v}{T})_{N,\kappa}$ for $\eta=\half{1}$ smaller than those found in the thermodynamic limit in \eqref{diffcvlim} -- as presented from numerical calculations in Fig. \ref{fig:dcv}. This is done by calculating $\qty(\pdv{c_v}{T})_{N,\kappa}$, finding its minimum in a large $N$ approximation\footnote{The calculation presented in this appendix was done in collaboration with D. Robbins.}.

Two variables will be important for this calculation. The first,  $\beta^*$, is the argument to the minimum value of the quantity of interest --- abscissa of the minimum values for each $N$ in Fig. \ref{fig:dcv} ---  meaning, $\beta^* \doteq \arg \min\limits_\beta \frac{1}{k_B^2  \beta} \qty(\pdv{c_v}{T})_{N,\kappa}$. The second, $\xi^*$ is defined as the fugacity at the minimum value of the quantity of interest, meaning $\xi^* \doteq \xi(\beta^*,N)$. From this, two other variables can be constructed: the reduced inverse temperature at the minimum $\gamma^* \doteq \frac{\beta^* -\beta_c}{\beta_c}$, and $\lambda_2^* \doteq - \log \xi^*$, whose notation $\lambda_2$ is inspired by it being the second Lagrange multiplier at the minimum, as explained in Appendix \ref{appendixMaxEnt}.

If one assumes a scaling relation between $\beta^*$ and $\lambda_2^*$  of the form 

\begin{equation}
    \label{scaling}
    \gamma^* = \Bar{\gamma} N^{-\psi} \ \qq{and} \ \lambda_2^* = \bar{\lambda}_2 N^{-\phi} \ ,
\end{equation}
in the leading order of $N$ --- where $\Bar{\gamma}$ {and} $\bar{\lambda}_2$ are constants and $\psi$ and $\phi$ are positive.
Substituting those variables in \eqref{LiN} and $\beta_c$ in \eqref{critical} one obtains  

\begin{equation}
    N = N (1+\gamma^*)^{-\half{3}} \frac{ \Li{\half{3}}(e^{-\lambda_2^*})}{\zeta(\half{3})} + \frac{1}{e^{\lambda_2^*}-1} \ .
\end{equation}
Note that accounting for only the first term would yield the regular  calculation in the thermodynamic limit --- expressed previously in \eqref{tq}.

Using the scaling relations in \eqref{scaling}, the series expansion for polylogarithms in \eqref{PLseries} and $e^{\lambda_2^*} = 1 +\lambda_2^* + o\qty(\lambda_2^*)$ one obtains, in the leading terms,

\begin{equation}\label{zero}
    0= -\frac{3}{2} \bar{\gamma} N^{1-\psi} - 2 \frac{\Gamma(\half 1)}{\zeta(\half{3})} \bar{\lambda}_2^{\ \half 1} N^{1-\half{\phi}} + \bar{\lambda}_2^{\ -1} N^{\phi} \ .
\end{equation}
A result that depends on both $\beta^*$ and $\lambda_2^*$  requires that the first and at least one other term in \eqref{zero} must contribute to the highest order in $N$. If only the first two terms contribute, the result would ignore the particles in the ground state, leading to the same results in the thermodynamic limit --- equivalent to \eqref{tq}. If only the first and last term in \eqref{zero} contribute, one would find $\bar{\gamma} = \frac{3}{2} \bar{\lambda}_2^{-1} $. This result however is undesirable physically, as observed in Fig. \ref{fig:dcv} we can expect $\beta^* < \beta_c$ and, consequentially, $\bar{\gamma} <0$; and for BE statistics one must have $\xi^* \leq 1$ implying $\bar{\lambda}_2 > 0$. 
Therefore, it follows that all terms in \eqref{zero} must contribute to the highest order, accounting for these terms one obtains $1-\psi = 1 - \half{\phi} = \phi$, hence $\psi = \third{1}$ and $\phi = \third{2}$  --- later these values will be verified numerically.

In order to obtain the values of $z(N)$ one needs to substitute $\beta^*$ and $\xi^*$ in \eqref{dcvdtob}. In order to do so, it is necessary to first substitute these values in $\frac{1}{\xi}\qty(\pdv{\xi}{\beta})_{N,\kappa}$ in \eqref{dxidb} and $\frac{1}{\xi}\qty(\pdv[2]{\xi}{\beta})_{N,\kappa} $ in \eqref{dxi2db2} as a expansion in terms of $N$. The parameters $\bar{\gamma}$ and $\bar{\lambda}_2$ will later be identified by imposing $\pdv{}{\beta} \qty[\frac{1}{k_B^2  \beta} \qty(\pdv{c_v}{T})_{N,\kappa}]\at{\beta=\beta^*} = 0$. These will be done in the following subsections.

\subsection{Expanding $\frac{1}{\xi}\qty(\pdv{\xi}{\beta})_{N,\kappa}$} \label{dxisec}
\paragraph{} One can expand the numerator of \eqref{dxidb}, $Q_{n}$ as

\begin{equation}
    Q_n = \frac{1}{\beta_c} \frac{3}{2} N + o\qty(N) \ ,
\end{equation}
where $o$ stands for the smaller order notation, $\lim\limits_{N\to\infty} \frac{o\qty(f(N))}{f(N)} = 0$.
Similarly, for the denominator of \eqref{dxidb}, $Q_d$, is expanded as

\begin{equation}\label{denominator}
    Q_d = a N^{\third{4}}(1+bN^{-\third{1}}) + o (N) \ ,
\end{equation}
where 

\begin{equation}\label{aandb}
    a = \frac{1}{\bar{\lambda}_2^{\ 2}} + \frac{\Gamma(\half{1})}{\zeta(\half 1)}\frac{1}{\bar{\lambda}_2^{\ \half{1}}} \ \qq{and} b = \frac{1}{a} \qty[ \frac{\zeta(\half{1})}{\zeta(\half{3})} - \frac{3}{2} \frac{\Gamma(\half{1})}{\zeta(\half{3})} \frac{\bar{\gamma}}{\bar{\lambda}_2^{\ \half{1}}} ]  \ .
\end{equation}
Therefore, using $\frac{1}{\xi}\qty(\pdv{\xi}{\beta})_{N,\kappa} = \frac{Q_n}{Q_d}$ it follows that

\begin{equation}\label{dxidb-appendix}
    \frac{1}{\xi}\qty(\pdv{\xi}{\beta})_{N,\kappa}\at{\beta=\beta^*} = \frac{1}{\beta_c} \bar{q} N^{-\third{1}} + o\qty(  N^{-\third{1}} ) \ ,
\end{equation}
where 

\begin{equation}\label{barq}
    \bar{q} = \frac{3}{2a} \ .
\end{equation}
{Note that the second term for $Q_d$ in \eqref{denominator} does not appear in \eqref{dxidb-appendix}. The importance of calculating the second term in $Q_d$ will be shown to be relevant when other quantities are calculated from it, as it will be done in the following subsection. }

\subsection{Expanding $\frac{1}{\xi}\qty(\pdv[2]{\xi}{\beta})_{N,\kappa}$} \label{d2xisec}
\paragraph{}
One can expand the numerator of \eqref{dxi2db2}, $F_{n}$ as

\begin{equation}
    F_n =  -\frac{a}{\beta_c^2} r_m N^{\third{4}}     + \frac{a}{\beta_c^2}  \bar{r} N + o\qty(N) \end{equation}
where

\begin{subequations}\label{rmandbarr}\begin{align}
    &r_m =   -\frac{1}{a^3} \qty[ \frac{9}{2} \frac{1}{\bar{\lambda}_2^3} + \frac{3}{2} \frac{\Gamma(\half{5})}{\zeta(\half{3})  } \frac{1}{\bar{\lambda}_2^{\ \half{3}}}  ]    \qq{and} \label{rm} \\ 
    &\begin{array}{ll} \bar{r} =  \dfrac{1}{a^3}&\left[ 
    -\dfrac{15}{4}a^2 + \dfrac{9}{2} \dfrac{\Gamma(\half{1})}{\zeta(\half{3})  } \dfrac{a}{\bar{\lambda}_2^{\ \half{1}}}  + \dfrac{9}{2} \dfrac{\Gamma(\half{5})}{\zeta(\half{3})  } \dfrac{b}{\bar{\lambda}_2^{\ \half{3}}}   + \dfrac{27}{2}  \dfrac{b}{\bar{\lambda}_2^{\ 3}} + \dfrac{39}{4} \dfrac{\Gamma(\half{5})}{\zeta(\half{3})  } \dfrac{\bar{\gamma}}{\bar{\lambda}_2^{\ \half{3}}} \right. \\ 
    & \left. \ \  + \dfrac{45}{2}  \dfrac{\bar{\gamma}}{\bar{\lambda}_2^{\ {3}}}   - {9} \dfrac{\Gamma(-\half{1})}{\zeta(\half{3})  } \dfrac{1}{\bar{\lambda}_2^{\ \half{5}}}   - 3 \dfrac{\Gamma(-\half{1}) \Gamma(\half{5}) }{(\zeta(\half{3}))^2  } \dfrac{1}{\bar{\lambda}_2}   
    \right] \label{barr} \ . \end{array}
\end{align}\end{subequations}
Note that the denominator of \eqref{dxi2db2} is the same as $Q_d$, expanded in \eqref{denominator}.
Therefore it follows that 

\begin{equation}\label{d2xidb2-appendix}
    \frac{1}{\xi}\qty(\pdv[2]{\xi}{\beta})_{N,\kappa}\at{\beta=\beta^*} = \frac{r_m}{\beta_c^2} + \frac{\bar{r}}{\beta_c^2}  N^{-\third{1}} + o\qty(  N^{-\third{1}} ) \ .
\end{equation}

\subsection{Expanding $\frac{1}{k_B^2  \beta} \qty(\pdv{c_v}{T})_{N,\kappa}$}
\paragraph{}
The quantity of interest, $\frac{1}{k_B^2  \beta} \qty(\pdv{c_v}{T})_{N,\kappa}$ in \eqref{dcvdtob}, can be expressed by substituting $\beta_c$ in \eqref{critical} as

\begin{equation}\begin{split} 
    \frac{1}{k_B^2  \beta} \qty(\pdv{c_v}{T})_{N,\kappa} = \frac{3}{2} \frac{1}{\zeta(\half{3})}   &
    \left[ \qty(\frac{\beta}{\beta_c})^{-\half 3} \ \frac{5}{2} \frac{3}{2} \ \Li{\half 5}(\xi)    \right. \\
    & \ \left. + \qty(\frac{\beta}{\beta_c})^{-\half 1}  \ \frac{\beta_c}{\xi}\qty(\pdv{\xi}{\beta})_{N,\kappa} \ \Li{\half 3}(\xi) \right. \\
    & \ \left. +  \qty(\frac{\beta}{\beta_c})^{\half 1} \ \qty(\frac{\beta_c}{\xi}\qty(\pdv{\xi}{\beta})_{N,\kappa})^2    \ \qty( \Li{\half1}(\xi) - \Li{\half3}(\xi) )\right. \\
    & \ \left.   +   \qty(\frac{\beta}{\beta_c})^{\half 1} \ \frac{\beta_c^2}{\xi}\qty(\pdv[2]{\xi}{\beta})_{N,\kappa}   \ \Li{\half 3}(\xi)    \right] \ .
\end{split}\end{equation}
using the expression for polylogarithms \eqref{PLseries} and the results of the previous subsections -- \eqref{dxidb-appendix} and \eqref{d2xidb2-appendix} --  this can be expanded as

\begin{equation}\begin{split} 
    \frac{1}{k_B^2  \beta} \qty(\pdv{c_v}{T})_{N,\kappa} \at{\beta=\beta^*} = z_m +\bar{z} N^{- \third{1}} + o\qty(N^{- \third{1}}) \ .
\end{split}\end{equation}
where 

\begin{subequations}\label{zmandbarz}
\begin{align} 
    &z_m =   \frac{3}{2}\frac{1}{\zeta(\half{3})} \qty[ r_m \zeta(\half{3})+ \frac{15}{4} \zeta(\half{5})]    \qq{and} \label{zm} \\ 
    &\begin{array}{ll} \bar{z} = \dfrac{3}{2}\dfrac{1}{\zeta(\half{3})}& \left[ \bar{q}^2 \Gamma(\half{1}) \bar{\lambda}_2^{\ -\half{1}}
    -3\bar{q} \zeta(\half{3})
    +\bar{r} \zeta(\half{3}) +\dfrac{1}{2} r_m \bar{\gamma} \zeta(\half{3})
     \right. \\ 
    & \left. \ \     
    +r_m \bar{\lambda}_2^{\ \half{1}}\Gamma(-\half{1})
    - \dfrac{45}{8}\bar{\gamma} \zeta(\half{5}) 
    \right] \label{barz} \ . \end{array}
\end{align}
\end{subequations}
Thus obtaining the scaling of $\frac{1}{k_B^2  \beta} \qty(\pdv{c_v}{T})_{N,\kappa}$ expressed in \eqref{zed}. In order to complete the goals of this appendix, one needs to obtain the values of $\bar{\gamma},\ \bar{\lambda}_2,\ a,\ b,\ \bar{q},\ r_m$, and $\bar{r}$ and substitute those in \eqref{zmandbarz}. This will be done in the next two subsections.

\subsection{ Obtaining $\bar{\gamma}$ and $\bar{\lambda}_2$}
\paragraph{}
Substituting the values of $\psi$ and $\phi$ in \eqref{zero}, it follows that

\begin{equation}\label{zero2}
   \frac{3}{2} \bar{\gamma}  + 2 \frac{\Gamma(\half 1)}{\zeta(\half{3})} \bar{\lambda}_2^{\ \half 1} - \bar{\lambda}_2^{\ -1} =0 \ .
\end{equation}
by implicit derivation  of the equation above one finds
\begin{subequations}
\begin{align}
    \dv{\bar{\lambda}_2}{\bar{\gamma}} &= -\frac{3}{2} \qty( \bar{\lambda}_2^{\ -2}  + \frac{\Gamma(\half{1})}{\zeta(\half{3})} \bar{\lambda}_2^{\ -\half{1}}  )^{-1} \label{dl2dg} \ , \\
    \dv[2]{\bar{\lambda}_2}{\bar{\gamma}} &= -\frac{2}{3} \qty(\dv{\bar{\lambda}_2}{\bar{\gamma}})^3 \qty( 2\bar{\lambda}_2^{\ -3}  + \frac{\Gamma(\half{3})}{\zeta(\half{3})} \bar{\lambda}_2^{\ -\half{3}}  ) \ , \label{d2l2dg2}\\
    \dv[3]{\bar{\lambda}_2}{\bar{\gamma}} &= -\frac{4}{9} \qty(\dv{\bar{\lambda}_2}{\bar{\gamma}})^5 \qty( 6\bar{\lambda}_2^{\ -6}  + \frac{\Gamma(\half{5})}{\zeta(\half{3})} \bar{\lambda}_2^{\ -\half{9}}  ) \ . \label{d3l2dg3}
\end{align}
\end{subequations}
The minimum occurs when $\pdv{}{\beta} \qty[\frac{1}{k_B^2  \beta} \qty(\pdv{c_v}{T})_{N,\kappa}]\at{\beta=\beta^*} = 0$, that means

\begin{equation}\begin{split} \label{mincond}
    0 = \frac{3}{2} \frac{\beta_c^{-1}}{\zeta(\half{3})}  
    &
    \left[ - \qty(\frac{\beta}{\beta_c})^{-\half 5} \ \frac{5}{2} \qty(\frac{3}{2})^2 \ \Li{\half 5}(\xi) +  \qty(\frac{\beta}{\beta_c})^{-\half 3}    \ \frac{5}{2} \frac{3}{2} \ \frac{\beta_c}{\xi}\qty(\pdv{\xi}{\beta})_{N,\kappa} \ \Li{\half 3}(\xi)    \right. \\
    & \ \left. + \qty(\frac{\beta}{\beta_c})^{-\half 1}  \ \frac{5}{2} \ \qty(\frac{\beta_c}{\xi}\qty(\pdv{\xi}{\beta})_{N,\kappa})^2 \ (\Li{\half 3}(\xi) -\Li{\half 1}(\xi)) \right. \\
    & \ \left. - \qty(\frac{\beta}{\beta_c})^{-\half 1}  \ \frac{5}{2}  \ \frac{\beta_c^2}{\xi}\qty(\pdv[2]{\xi}{\beta})_{N,\kappa} \ \Li{\half 3}(\xi) \right. \\
    & \ \left. + \qty(\frac{\beta}{\beta_c})^{\half 1}  \  \qty(\frac{\beta_c}{\xi}\qty(\pdv{\xi}{\beta})_{N,\kappa})^3 \ (2\Li{\half 3}(\xi) -3\Li{\half 1}(\xi))+\Li{-\half 1}(\xi)) \right. \\
    & \ \left. - \qty(\frac{\beta}{\beta_c})^{\half 1}  \ 3 \ \qty(\frac{\beta_c}{\xi}\qty(\pdv{\xi}{\beta})_{N,\kappa}) \qty(\frac{\beta_c^2}{\xi}\qty(\pdv[2]{\xi}{\beta})_{N,\kappa}) \ (\Li{\half 3}(\xi) -\Li{\half 1}(\xi)) \right. \\
    & \ \left.   +   \qty(\frac{\beta}{\beta_c})^{\half 1} \ \frac{\beta_c^3}{\xi}\qty(\pdv[3]{\xi}{\beta})_{N,\kappa}   \ \Li{\half 3}(\xi)    \right] \ .
\end{split}\end{equation}
In order to solve \eqref{mincond} one may have to expand $\frac{1}{\xi}\qty(\pdv[3]{\xi}{\beta})_{N,\kappa} $ --- as done for $\frac{1}{\xi}\qty(\pdv{\xi}{\beta})_{N,\kappa} $ in Sec. \ref{dxisec} and $\frac{1}{\xi}\qty(\pdv[2]{\xi}{\beta})_{N,\kappa} $  in Sec. \ref{d2xisec}. 
However, a less laborious manner to perform this calculation involves identifying from \eqref{scaling} that

\begin{subequations}
\begin{align}
    \frac{\beta_c}{\xi}\qty(\pdv{\xi}{\beta})_{N,\kappa} = -\dv{\lambda_2^*}{\gamma^*} = & -\dv{\bar{\lambda}_2}{\bar{\gamma}} N^{-\third{1}} + o\qty( N^{-\third{1}})  \ , \label{d1superapprox}\\
    \frac{\beta_c^2}{\xi}\qty(\pdv[2]{\xi}{\beta})_{N,\kappa} = -\dv[2]{\lambda_2^*}{{\gamma^*}} = & -\dv[2]{\bar{\lambda}_2}{\bar{\gamma}}  + o\qty(1) \ , \qq{and} \label{d2superapprox}\\
    \frac{\beta_c^3}{\xi}\qty(\pdv[3]{\xi}{\beta})_{N,\kappa} = -\dv[3]{\lambda_2^*}{{\gamma^*}} = & -\dv[3]{\bar{\lambda}_2}{\bar{\gamma}} N^{\third{1}} + o\qty( N^{\third{1}})  \ .
\end{align}
\end{subequations}
Later it will be shown that $\bar{q} = -\dv{\bar{\lambda}_2}{\bar{\gamma}}$, as expected from \eqref{dxidb-appendix}, and $r_m = -\dv[2]{\bar{\lambda}_2}{\bar{\gamma}}$, as expected from \eqref{d2xidb2-appendix}\footnote{Note that this does not invalidate the work done in Sec. \ref{dxisec} and \ref{d2xisec}, since \eqref{d2superapprox} does not obtain the second term in \eqref{d2xidb2-appendix}.}. 

Expanding \eqref{zero2} it follows that 

\begin{equation}
    0 = -\dv[3]{\bar{\lambda}_2}{\bar{\gamma}} N^{\third{1}} \zeta(\half{3})+ o\qty( N^{\third{1}})  \ .
\end{equation}
Hence, the minimum condition implies $\dv[3]{\bar{\lambda}_2}{\bar{\gamma}}=0$, which is equivalent, per \eqref{d3l2dg3}, to 

\begin{equation}
    6\bar{\lambda}_2^{\ -6}  + \frac{\Gamma(\half{5})}{\zeta(\half{3})} \bar{\lambda}_2^{\ -\half{9}}  = 0 \implies \bar{\lambda}_2 = \qty[\frac{1}{6} \frac{\Gamma(\half{5})}{\zeta(\half{3})} ]^{-\third{2}} \ ,
\end{equation}
and applying this into \eqref{zero2} it follows that

\begin{equation}
    \bar{\gamma} = \frac{2}{3} \qty[ \bar{\lambda}_2^{\ -1}-2\frac{\Gamma(1/2)}{\zeta(3/2)}\bar{\lambda}_2^{\ 1/2} ] \ ,
\end{equation}
leading to the values 
\begin{equation} \label{bars}
\bar{\lambda}_2 \approx 5.1804 \qq{and} \bar{\gamma} \approx -1.9303 \ .
\end{equation}

\begin{figure}
\centering
    \includegraphics[width=.4\textwidth]{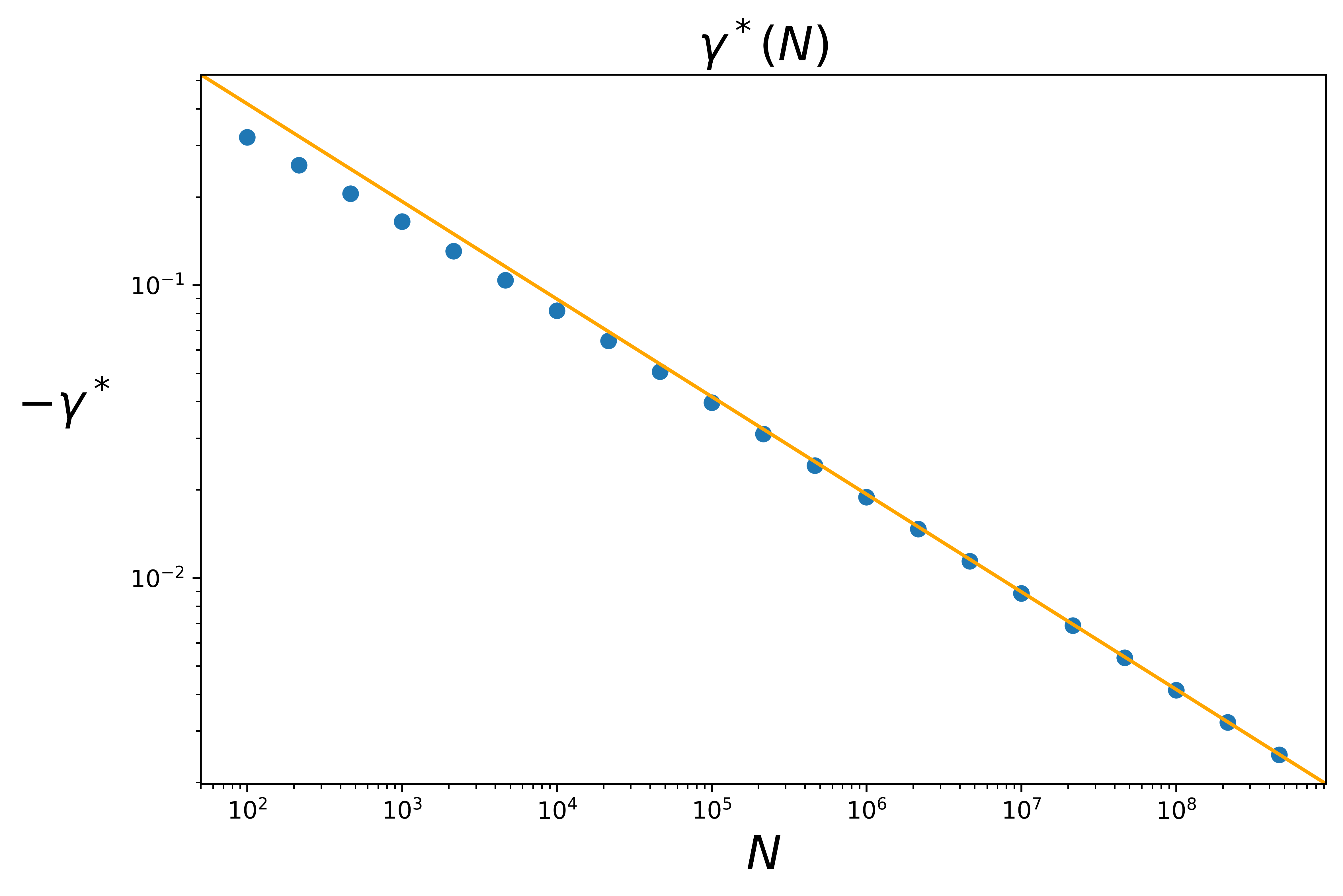}
    \includegraphics[width=.4\textwidth]{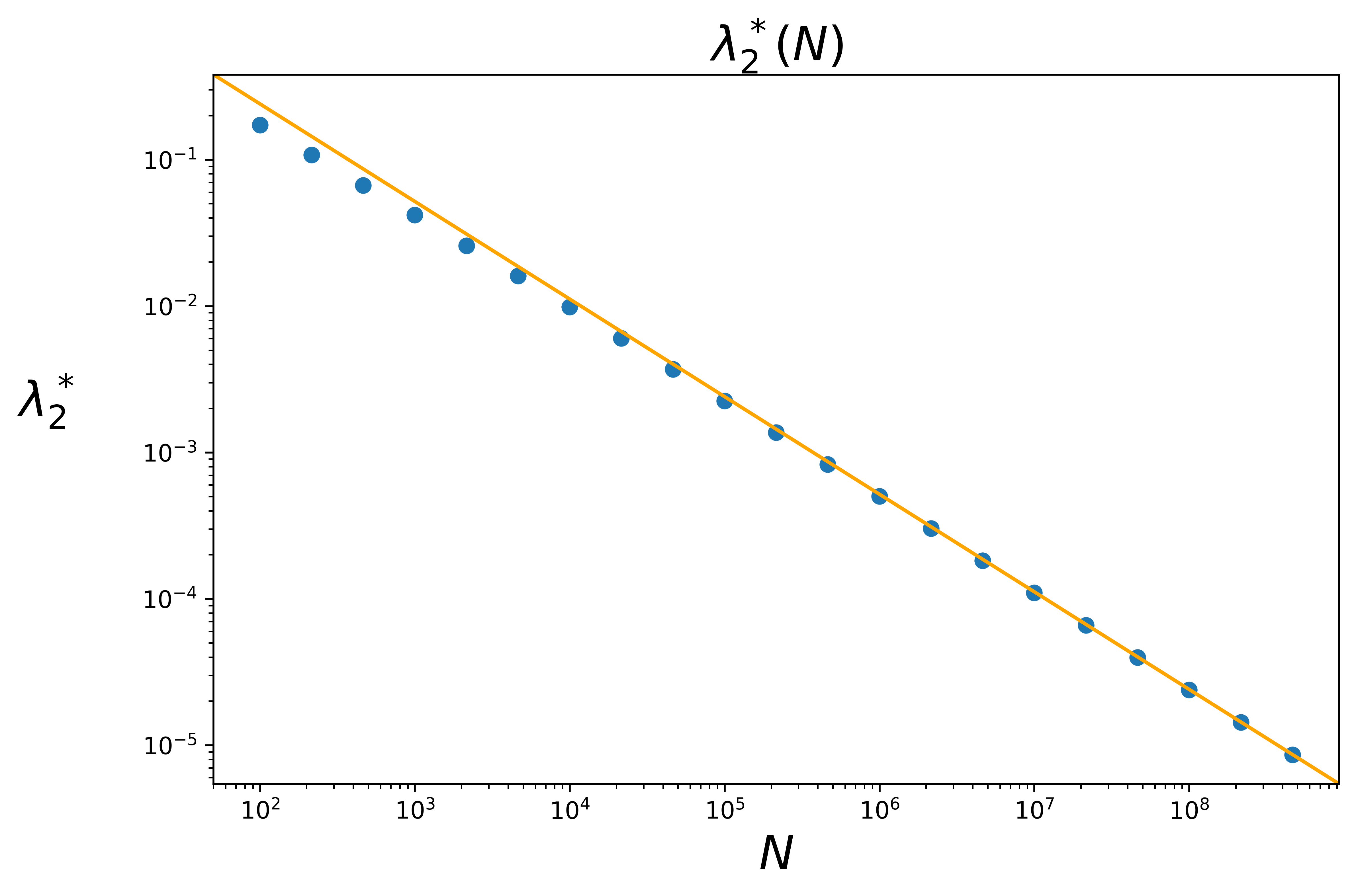}
    \caption{Graph for the values of $\gamma^*$ (left) and $\lambda_2^*$ (right) calculated numerically for $N$ ranging from $10^2$ to $10^8$ (scattered blue points) and the approximation in order of $N^{-\third{1}}$ and  $N^{-\third{2}}$ (solid orange line) respectively --- meaning $\gamma^*(N) =  \bar{\gamma} N^{-\third{1}}$ and  $\lambda_2^*(N) =  \bar{\lambda}_2 N^{-\third{2}}$ as in \eqref{scaling}, with $\bar{\lambda}_2$ and $\bar{\gamma} $ given by \eqref{bars}.  }
    \label{fig:gl2ed}
\end{figure}

The numerical verification of \eqref{scaling} with these values of $\bar{\gamma}$ {and}  $\bar{\lambda}_2$ is presented in Fig. \ref{fig:gl2ed}.

\subsection{Obtaining $a,b,\bar{q},r_m, \bar{r},z_m$, and $\bar{z}$}
\paragraph{}
Substituting the values of $\bar{\gamma}$ {and}  $\bar{\lambda}_2$ from \eqref{bars} into \eqref{aandb}, one obtains

\begin{equation} \label{finalab}
a \approx 0.33536 \qq{and} b \approx  0.90686 \ .
\end{equation}
Sequentially applying these values in \eqref{barq} yields 

\begin{equation} \label{finalbarq}
    \bar{q} \approx 4.47284 \ .
\end{equation}
Note that substituting the value of $\bar{\lambda}_2 $ from \eqref{bars} into \eqref{dl2dg} implies $\bar{q} = - \dv{\bar{\lambda}_2}{\bar{\gamma}}  $ in accordance to \eqref{d1superapprox}. Similarly, applying \eqref{finalbarq}, \eqref{finalab}, and \eqref{bars} into \eqref{rmandbarr} yields

\begin{equation} \label{finarmandlbarr}
    r_m \approx -2.5746 \qq{and}   \bar{r} \approx -6.1656 \ .
\end{equation}
Note that substituting the value of $\bar{\lambda}_2 $ from \eqref{bars} into \eqref{d2l2dg2} implies $r_m = - \dv[2]{\bar{\lambda}_2}{\bar{\gamma}}  $ in accordance to \eqref{d2superapprox}.  
Finally, substituting \eqref{bars}, \eqref{finalbarq}, and \eqref{finarmandlbarr} into \eqref{zmandbarz} one obtains
$z_m$ and $\bar{z}$ as in \eqref{zedvalues} completing the calculation. 

\vspace{.5cm}
\bibliographystyle{naturemag-doi}
\bibliography{ref}


\end{document}